\documentclass[acmsmall]{acmart}
\usepackage{multirow}
\usepackage{longtable}
\usepackage{array}
\AtBeginDocument{%
  \providecommand\BibTeX{{%
    \normalfont B\kern-0.5em{\scshape i\kern-0.25em b}\kern-0.8em\TeX}}}

\setcopyright{acmlicensed}
\acmJournal{PACMHCI}
\acmYear{2023} \acmVolume{7} \acmNumber{CSCW1} \acmArticle{29} \acmMonth{4} \acmPrice{15.00}\acmDOI{10.1145/3579462}

\begin{document}
\title[Beyond Transactional Democracy]{Beyond Transactional Democracy: A Study of Civic Tech in Canada}

\author{Curtis W McCord}
\email{curtis.mccord@mail.utoronto.ca}
%\orcid{Placeholder}
\affiliation{%
\institution{University of Toronto}
   \streetaddress{140 St. George St.}
   \city{Toronto}
   \state{Ontario}
   \country{Canada}
   \postcode{M5S 3G6}
 }
 \author{Christoph Becker}
 \affiliation{%
   \institution{University of Toronto}
   \streetaddress{140 St. George St.}
   \city{Toronto}
   \country{Canada}}
 \email{christoph.becker@utoronto.ca}

\renewcommand{\shortauthors}{McCord and Becker}

\begin{abstract}
Technologies are increasingly enrolled in projects to involve civilians in the work of policy-making, often under the label of `civic technology'. But conventional forms of participation through transactions such as voting provide limited opportunities for engagement. In response, some civic tech groups organize around issues of shared concern to explore new forms of democratic technologies. How does their work affect the relationship between publics and public servants?

This paper explores how a Civic Tech Toronto creates a platform for civic engagement through the maintenance of an autonomous community for civic engagement and participation that is casual, social, nonpartisan, experimental, and flexible. Based on two years of action research, including community organizing, interviews, and observations, this paper shows how this grassroots civic tech group creates a civic platform that places a diverse range of participants in contact with the work of public servants, helping to build capacities and relationships that prepare both publics and public servants for the work of participatory democracy.  

The case shows that understanding civic tech requires a lens beyond the mere analysis or production of technical artifacts. As a practice for making technologies that is social and participatory, civic tech creates alternative modes of technology development and opportunities for experimentation and learning, and it can reconfigure the roles of democratic participants. 
\end{abstract}

\begin{CCSXML}
<ccs2012>
<concept>
<concept_id>10003120.10003130.10011762</concept_id>
<concept_desc>Human-centered computing~Empirical studies in collaborative and social computing</concept_desc>
<concept_significance>500</concept_significance>
</concept>
</ccs2012>
\end{CCSXML}

\ccsdesc[500]{Human-centered computing~Empirical studies in collaborative and social computing}

\keywords{civic technology, participatory democracy, digital civics, civic engagement, citizen participation}

\maketitle

\section{Introduction}
A core assumption of democratic theory is that for democracy to work, there must be public discourse that checks and legitimates government actions. This is part of what what J\"{u}rgen Habermas called the ``public sphere'' and what we take to refer to the largely free, interested, and voluntary discourse that citizens or civilians (we use these terms interchangeably), actors outside of representative or public service roles undertake in relation to state representative or institutional actions (or inaction)~\cite{habermas1989structural}. For a long time, the primary role of a public sphere was discursive, but there are wider roles for participation within policy-making today. In part, this is because new technologies have not only produced new kinds of objects for engagement (such as software systems and data sets) but also new ways of structuring engagements. 

The process of public technology design creates additional responsibilities for public servants who have public-facing roles in facilitating these processes, while participating civilians need additional knowledge of technological artifacts and their development processes. To contribute effectively to these emerging democratic processes, people require literacies and supports that befit these new objects. However, while technology has changed how political participation can occur, it is not clear that it has meaningfully expanded the role or agency of participants in democratic governance. Top-down modes of public consultation, even when they are technologically sophisticated, often exemplify individualistic and transactional approaches to engagement, where inputs are solicited in isolation and handled within a black box of decision-making~\cite{vlachokyriakos2016digital,mcdonald2019}. This suggests an opportunity to examine how self-organized groups of people participate in the work of policy-making.  

Some groups have organized around issues of shared concern to explore new forms of democratic technologies \cite{chock2020design,g0v2022g0v,ctpoland2022cto,ajl2022home,garcia2021dci}. They exemplify how community coalitions can build the capacities required for sociotechnical citizenship, and even seek out engagements with governance. In this paper, a \textit{civic tech} group in Canada serves as an example of how a diverse group of participants, including civilians and public servants, form communities that revolve around civic engagement and a commitment to interacting with governments. Documenting the role and contributions of these relationships not only addresses the epistemic contribution of participatory processes to decision making, but also provides a suitable basis to consider how civic engagement around issues of technology design, policy and discourse, can make policy and design processes more legitimate, visible, and even participatory. 

What is the role of technology design in democratic life beyond the automation of existing democratic transactions, such as voting? Who can, will, and could participate in it? This paper explores how, through the maintenance of an autonomous community for civic engagement and participation that is casual, social, nonpartisan, experimental, and flexible: Civic Tech Toronto\footnote{For purposes of peer review, names have been pseudonomized; in final publication, the real names of these organizations, places, and people will appear in the text} (CTTO), which creates a platform for civic engagement. By mediating collaboration and interaction between civilians and public sector institutions, building capacities in civil servants and laypeople, and hosting projects, CTTO has begun to create an infrastructure for participation and political expression whose potential value is clear, but unrealized. 

To begin, we outline frameworks for analyzing participatory processes and techniques and review how civic action and participation are discussed in CSCW and related fields (Section~\ref{sec:background}). Following an account of the research methodology and site  (Section~\ref{sec:method}), we provide an account of CTTO (Section~\ref{sec:CTTO}). It shows how this civic tech group, operating as a platform, provides settings and opportunities for participation that are distinct from, and complementary to, the types of engagement typically available to states. Section~\ref{sec:discuss} situates these arguments within the field while in Section~\ref{conc} we conclude by discussing potential research directions.

\section{Background} % Revisions here
\label{sec:background}
Careful study of democratic participation requires attention to the way relationships between people and power are envisioned and enacted. In any process intended to involve people into the institutional work processes of policy-making or design, the roles assigned to civic actors within the process will reflect value-laden assumptions about what people can or ought to contribute to these typically technical and political exercises. Below, we cover some of the important concepts from our argument: \textit{civic engagement}, \textit{participation}, \textit{publics}, and \textit{civic tech}. We stabilize these terms and examine how these issues are seen within CSCW and related fields.

\subsection{Civic Engagement}
The distinguishing characteristic of democracies is that claims to the legitimacy of the regime rest upon consent of the governed. To do this, systems must be created that credibly translate popular intention and will into state or collective action, such as voting and representation. The relationships between governing and governed are sociotechnical, as much about the arrangement of institutional affordances of technologies as they are about the social norms and practises of a society, and are thus configurable~\cite{coleman2012making}. 

Political subjectivity, or the ways that people experience or imagine or act within or alongside political systems (which are themselves constructed), can be construed in different ways. As a way of specifying political subjectivity, the same is true of citizenship, one of the more commons ways of denoting political subjects. Some conceptions of citizenship make a clear distinction between public and private life and see the public sphere and the state as a guarantor for the freedoms of the private sphere. Here, citizenship is primarily a legal category, exercised only occasionally (e.g. in voting, or in court), which guarantees rights, entails responsibilities, and gives access to services that allow the pursuit of goals in private life~\cite{leydet2014citizenship}. 
Other conceptions of democratic citizenship, or of civic life, are based on the idea that political agency can only be realized through active participation, either in the decision-making processes of government or through collective action with others~\cite{leydet2014citizenship}. We call these activities, both participation in state-sanctioned decision-making processes and collective action to address shared matters of concern, \textit{civic engagement}. At its core, this stronger view of civic life expresses an often expressed principle of democracy, from Rousseau ~\cite{rousseau1997social}, to Critical Systems Heuristics ~\cite{ulrich1983critical}, to Participatory design ~\cite{harrington2019deconstructing}; that those affected by decisions should have a place in making them.

Today, democratic states are developing more participatory mechanisms for decision-making not only to realize practical goals of taking in more information and perspectives from those they represent, but in order to legitimate their decision-making amidst a time of increasing disaffection and alienation. Networking and information technologies, offering vast increases of scale and speed of communication, are seen and treated as available medium for increasing participatory opportunities. Meanwhile, technologies and firms are the object of policy interventions by government and civil society ~\cite{crabu2018bottom}. Of course, there are major issues in realizing deeply participatory modes of democracy. Questions of pluralism aside, the sheer complexity of state institutions makes entry into decision-making technically burdensome not only for people, but for the institutions that would coordinate and make sense of this engagement. 

There is significant scholarly interest in government use of ICTs to support and deepen civic engagement \cite{Medaglia2012,saldivar2019civic,zhang2021review}. These `civic' technologies play an important role in state and civil society efforts to increase meaningful democracy in our societies. However, this paper is not focused on these technologies. Rather, we focus on  civic technology as part of a social world of grassroots initiatives and community based civic technology development that engages with the state. To this end, we provide a background that emphasizes the social elements of civic technology and examine how this view troubles perspectives that are too narrowly focused on technological artefacts. A review of the general use and design of technological innovations in civic contexts, or of social media and ICT use by governments, is out of scope.

\subsection{Participation and Pseudo-Participation}
If civic engagement configures a relationship between the governed, the government, or governance, how can we make sense of it? One way is through examining the roles that people play in decision-making processes. Expanding on Arnstein's ``Ladder of Participation''~\cite{arnstein1969ladder}, Cardullo and Kitchin developed a ranked classification of types of ``smart citizen'' participation in smart city projects~\cite{cardullo2019being}. At the core of this ladder are questions about the distribution of decision-making power in engagement processes and about the relationship between expertise, power, and the agency of those affected by decisions. In the democratic ideal expressed by this ladder, those affected by decisions have a genuine say in their making. At the bottom rungs of the ladder are types of non-participation, where communication is one-way and citizens are immanent to the decisions of the state. As we climb the ladder, we encounter forms of consumerism and tokenism, where contributions are sought but their influence is fully mediated by the incumbent decision-makers, be they planners, politicians, etc. That is, while those affected are invited to participate, they do so only on grounds already circumscribed by those controlling the process. The actual decision-making power of a system under design remains unchallenged, cf. ~\cite{ulrich1983critical,Ulrich2010}. This disconnect between participation and control of a decision-making process has also been called ``pseudo-participation'', a form of ``participation without agency'' ~\cite[p.41]{palacin2020design}. For Palacin et al., pseudo-participation \textit{in design} occurs when participation in a decision making process is shaped to sharply constrain what participants can do, for example in crowd-sourcing or public naming competitions. Pseudo-participation \textit{by design} occurs when these constraints on participation are embedded into design artifacts that control a process. In both cases, the process itself is not subject to participatory decision-making and the degree of freedom of participants is severely constrained. 

There are potential weaknesses in Cardullo and Kitchin's ladder. At the top of the ladder is citizen power, where citizens have or share authority to make decisions, contributing ideas, visions, and leadership to the process~\cite{cardullo2019being}. Civic hacking and hackathons are presented as some of the most participatory and empowering forms of engagement. However, there are obstacles to realizing these potentials, and civic hacking and hackathons are still vulnerable to the same marginalization as other participatory processes. For example, Johnson and Robinson note that without an account of outcomes, hackathon production could be performative in nature if the products of these events are not implemented and maintained long-term ~\cite{johnson2014civic}. Additionally, they are cautious about how framing control and compensation effect the distribution of power in these hackathons; are participants sufficiently empowered within these processes to influence the priorities for the project, or are they merely contributing free labour? In a one-off sprint, we may worry that the hackathon format privileges solutionist responses based on oversimplifications of what might be complex social problems, as Lilly Irani finds~\cite{irani2015hackathons}. Additionally, unless supported by a reflexive and inclusive methodology, hackathons do little to involve those already marginalized within public discourse; this seems particularly worrisome when participants are solicited on the basis of technical (usually programming or development) expertise, rather than as a public connected to an issue~\cite{irani2015hackathons,chock2020design,rosner2018critical}.  

\subsection{Publics and Communities}
In fields like HCI and CSCW, \textit{publics} play an important role in describing group formation and collective action, placing design and making into an account of public life. Tracing its origins to the work of John Dewey, the concept of a public is used to denote a set of people who by stake, interest, and performance are connected around a shared \textit{issue} ~\cite{dewey2012public}. Issues and publics are co-constituted discursively: a public is formed by a shared articulation of an issue and a set of actions taken to address it ~\cite{disalvo2014making}. As such, the boundaries of publics are porous. Membership is non-exclusive and based on participation within the discourse and action around an issue. 

The concept of \textit{a public} is an important correction to the notion of a monolithic general public~\cite{dewey2012public,disalvo2016designing}. Drawing on Dewey's formulation, Le Dantec provides more concepts that can help us analyze the performance of publics. Publics form around the negotiation and stabilization of an issue, but also by other means such as interventions and \textit{attachments}, i.e. their commitments to, and actions in service of, their issues. By connecting actors, artifacts and institutions, attachments create new possibilities for action and lead to the creation of durable systems and relationships that afford future action around issues ~\cite{le2016designing}. As publics concretize around their issues, they are either engaged with objects that are themselves designed, or their actions take the form of design. 

This process, known as \textit{infrastructuring}, deepens the involvement of publics in action, and even decision-making, when it can be connected to power ~\cite{le2016designing}. For example, Weise et al. describe how planners and, by extension, public servants embedded in engagement processes, play a crucial role as ``infrastructuring agents\ldots{} who are meant to enable civic groups'' ~\cite[p.537]{weise2020infrastructuring}. In this capacity, planners must cultivate relationships with publics that can productively support their work. This extends also to using collaborative and iterative methods in designing the engagement process itself. This focus on the importance of infrastructuring work helps to shift the focus of engagement from exchanges to relationships, thereby decentering technologies and including their context of use. Infrastructuring thus creates ``a continuous alignment between different communities'' ~\cite[p.104]{marttila2017infrastructuring}. This gives some clarity to the work of civic engagement: it is perhaps problematically ambitious to demand that states should engage ``the public'', but states \textit{can} cultivate relationships with specific publics to further the goals of participatory democracy by drawing boundaries in terms of issues and attachments.

\subsection{Civic Tech and Digital Civics}
At a 2020 workshop at the CSCW conference, participants discussed the research agenda for civic technologies. The workshop proposal notes that studies on civic technologies have prioritized ``technological innovation rather than social impact\ldots{} [and] results \ldots{} [rather than] community practise'' ~\cite{aragon2020civic}. At least partially, this gap in research seems related to the way that civic technology is treated as a kind of technology, rather than as a process for designing socio-technical systems. For example, in a systematic literature review, Saldivar et al. define ``civic technology\dots[as] technology (mainly information technology) that facilitates democratic governance among citizens'' ~\cite[p.170]{saldivar2019civic}. In some cases, this focus has privileged a focus on technologies deployed by governments, similar to terms like ``public interest technology''~\cite{Schneier2019} (technology services aimed at managing public goods and deploying public resources), ``eGovernment'' and ``digital government'' (networked information systems that provide access to government services) and ``eParticipation'' (technologies that enable online interactions with democratic processes) ~\cite{Macintosh2004,Medaglia2012}. Saldivar et al. are sensitive to the shortcomings of a transactional focus on engagement and highlight that prior CSCW research is limited by it~\cite{saldivar2019civic}. 

In some cases, researchers handle civic engagement explicitly within a transactional frame, for example, examining crowdsourcing as civic engagement, but with the qualification that the social context of these technologies is a much better explanation for their success ~\cite{clark2019citizen}. Ludwig et al. explore the emergence of publics within social networking technologies. For them, engagement happens when ``an individual undertakes an action on a mobile device with regard to the current situation''~\cite[p.211]{ludwig2016publics}. Considering a crowdsourcing system deployed by the Guardian newspaper to sort through over two million documents related to the expenses of British members of parliament, Handler and Conill treat civic technologies as those ``speciﬁcally created to enable, facilitate, and enact civic participation'', but are frank in admitting that, ``the interface alone would not have resulted in civic participation\ldots{}  game mechanics allowed people to access open data they had an interest in analysing''~\cite[p.161]{handler2016open}. In other words, the technology alone could not explain the popularity of the exercise; there already existed an interest in exposing potential corruption, and the stewardship of a major media outfit meant that the work of contributors could reasonably be seen as leading to results (i.e. press coverage). 

By contrast, work by scholars such as Lilly Irani and Chris le Dantec is much more focused on the social practises and processes of civic engagement within the context of ``entrepreneurial citizenship'' (where states promote market-based innovation culture as a public good) ~\cite{Irani2019,irani2015hackathons} and  ``social design'' (where publics engage in design processes to better their shared situations) ~\cite{le2016designing}. 

While we can understand a lot by examining discrete instances of public participation, research in this space must also develop ways of situating these processes within the larger contexts of democracy and seek out ways to examine how these practises of design can lead to durable infrastructures that increase the effectiveness and availability to participatory methods within institutional and community contexts. In this vein, Vlachokyriakos et al. offer the concept of ``\textit{digital civics}'' and provide a critique of transactionalist conceptions of citizenship, noting that: ``contemporary public service provision casts citizens as service consumers (even customers)\ldots{} [or as] objects of an issue, and as responsible for feeding back on services without a genuine involvement in the shaping of the service provided.''~\cite{vlachokyriakos2016digital}. This is echoed by McDonald's study of the US congress, which finds that the technology employed to interact with citizens ``creates discursive expectations that prioritize data-collection and customer satisfaction over substantive engagement''~\cite[p.17]{mcdonald2019}.

Digital civics encourages the ongoing collaboration between local authorities and citizens to ``create a participatory imaginary\ldots{}[to] explore the value of relational models of service provision'' ~\cite[p.1098]{vlachokyriakos2016digital}. While still generally focused on the ways that technologies might be the primary mediators of these potential new modes of governance, for Vlachokyriakos et al., the substantive problem for civic engagement is to re-configure the role of citizens within the work of governance itself, and to reconfigure the role of public servants in facilitating this work, by bringing the two into closer proximity. 

This reconfiguration of roles and relationships is at the core of the practise that we will call \textit{civic tech}. Civic tech is a view of ``government as platform'' rather than as a ``vending machine'', alluding to a larger vision of how technology and democratic governance can be made compatible ~\cite{schrock2018civic}. As a practise, civic tech is inherently experimental, seeking new ways to bring new communities and expertises into the work of policy-making~\cite{whitaker2015what,peer2019designing}, realizing goals of ``building with, not for''~\cite{McCann2015}, and heralding new a methodological paradigm for public servants~\cite{harrell2020civic}. In brief, civic tech is about realizing that technologies and processes of technology development are civic issues and about finding ways to make decisions around them more democratic and inclusive. 

\subsection{Summary and Research Directions}

The themes of civic agency and participation appear throughout these conversations and suggest fruitful perspectives. As Dickinson et al. write, ``[t]he transactional, data-driven approach to designing civic tech does not support relationship-building between city governments and communities and does not leverage (or even acknowledge) existing community assets''~\cite{dickinson2019cavalry}.  This suggests the need to look beyond transactional models and focus on community relationships. Because ``digital civics focuses on the relational elements comprising the fundamentals of civic interaction''~\cite{asad2017creating}, studying it calls for prolonged, situated engagement and long-term interactions with communities~\cite{erete2017empowered,joshi2021flaky,prost2018food,dickinson2019cavalry,Irani2019,le2016designing}.

Building community participation takes time. Civilians need mediating entities that prepare and empower them to enact democratic roles and to effectively organize their participation in democratic processes. Studies of technologies for engagement and studies of the involvement of participants of these processes need to examine how these relationships can be built over time, why they struggle to emerge, and what sorts of entities and practises might equip both civil and state actors to engage in a more social and participatory manner.

Civic tech as a term captures not only sets of technologies, but \textit{practises} that use and design  technological artifacts to create speculative relationships between civilians and states. Relational approaches to the study of civic technology require that we de-centre the technological artifice of civic tech and explore how these practises imagine and infrastructure relationships between people and states. Accordingly, a sociotechnical approach to civics locates political agency within practises of civic engagement that create relationships between people and between people and states. This area of political performance and participation, which we call civic life, is configured through social practise, technological mediation, and institutional affordance. 

Central current issues in the study of civic tech include how the ``local context and infrastructure affect the design, implementation, adoption, and maintenance of civic technology''; how civic technology development connects government, civilian and organizational actors, how it builds trust, and how to study such phenomena~\cite[539f]{aragon2020civic}. According to Asad et al.~\cite{asad2017tap}, ``current scholarship has overlooked some of the complexities in civic work'', and more attention should be paid to the issue-advocacy work that often spans multiple sites.  Joshi et al. call to couple such ``empirical analyses of citizenship'' with ``more heterogeneous conception of public service infrastructures''~\cite{joshi2021flaky}.  

In this paper, we contribute to this emerging interest in the sociotechnical relationships of civic technology and digital civics using an example from the Canadian civic tech community. Through extended community based research including weekly ``hacknights'' organized around the discussion and design of technologies within a shared social space, we explore how civic tech can contribute to civic life and create positive and productive civic relationships. Specifically we address two questions:
\begin{enumerate}
\item How does civic tech connect government and civilian actors?
\item How does civic tech practice reconfigure the roles and relationships of participating actors?
\end{enumerate}

\section{Methods and Study Design}\label{sec:method}
\subsection{Action Research}

\begin{figure}
    \centering
    \includegraphics[scale=0.45]{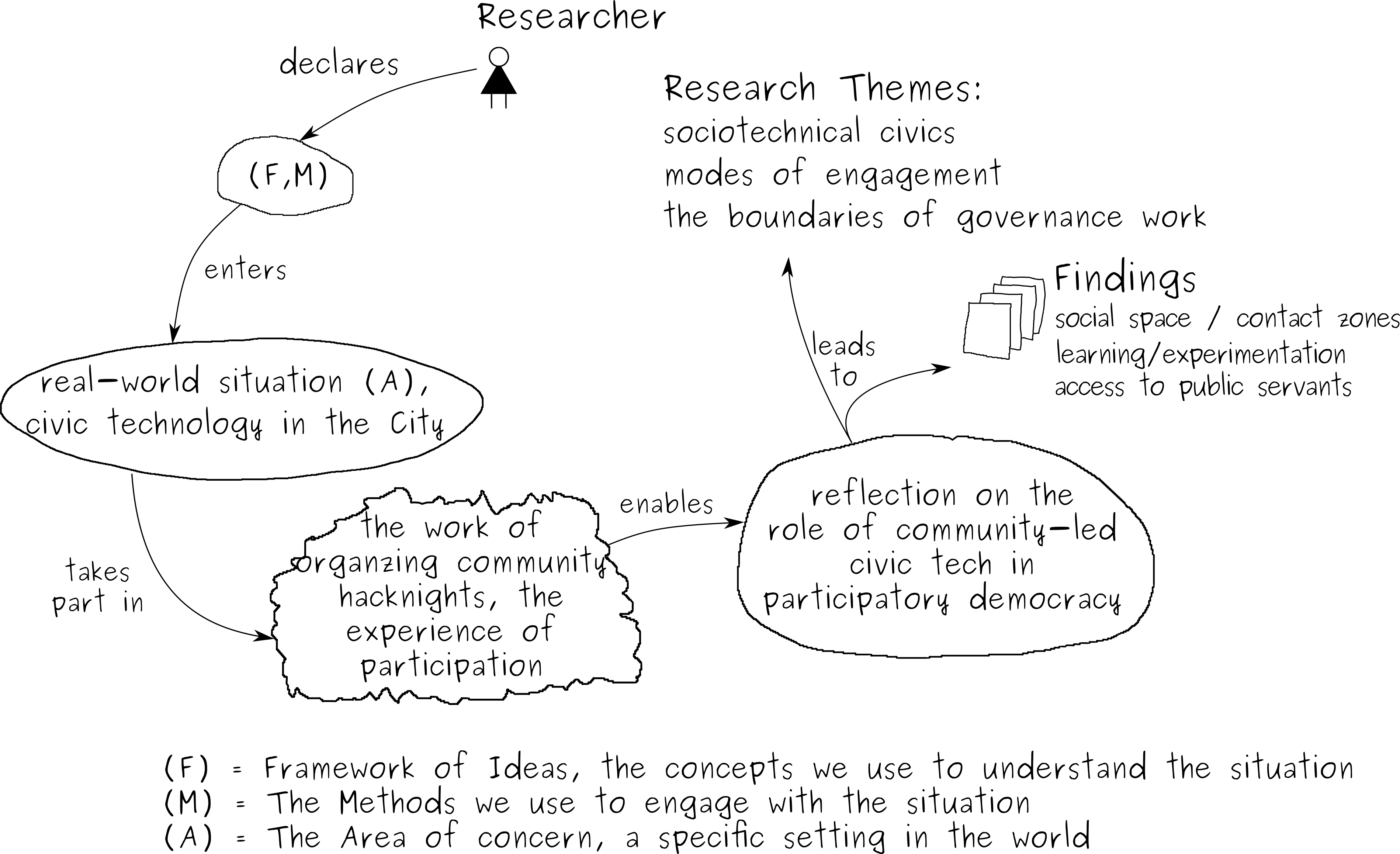}
    \caption{An adaptation of Checkland and Howell's~\cite{checkland1998action} diagram summarizing the process of action research within Soft Systems Methodology (SSM), adapted to this project}
    \label{ar-diag}
\end{figure}

This research was structured as an action research project~\cite{hayes2011relationship} in line with Checkland and Howell's methodology~\cite{checkland1998action}, wherein participation in a situation is used to elicit research themes while also contributing actively to the day-to day-work of the organization (see Figure\ref{ar-diag}). This framework for action research is closely aligned with those others employ in HCI research~\cite{Irani2019,joshi2021flaky,hayes_knowing_2014} in terms of the iteration between reflection and intervention~\cite{prost2018food} and in terms of the kinds of research motivations and claims that it makes~\cite{hayes2011relationship}. 

In SSM-inspired action research, the researcher declares a set of frameworks and methods that will motivate their work in a real-world situation. Since we were interested in exploring the relationships, infrastructures and objects that structure the civic performances of CTTO, and how CTTO related to other entities such as firms, nonprofits, and governments the Social Worlds/Arenas Framework (SWA/F) is particularly appropriate~\cite{clarke2008social} and  helps to inform (along with action research) a set of methods and analytic techniques. Naming a social world draws attention to groups of people organizing around common sets of activities using complex technologies, sharing specialized language in communication~\cite{Strauss1978}. These social worlds can intersect, separate, and meet in overarching social arenas~\cite{Strauss1982}. 

SW/AF includes its own framework for analysis (Situational Analysis, discussed in more detail in Subsection~\ref{anal}) and the language of Social Worlds. In addition, it encourages researchers to direct their inquiry and analysis using ``sensitizing concepts [which] suggest directions along which to look'' ~\cite[p.28]{clarke2005situational}. We drew on ideas of social production~\cite{Hardt2017} and social reproduction~\cite{caffentzis2014commons} to understand what and how production (especially the production of social relations) could be considered in the context of CTTO, and how the community reproduced itself. Interested in the relationship between civic technology production and the state, we also kept in mind concepts of citizenship, especially Irani's notion of ``entrepreneurial citizenship''~\cite{Irani2019}. Finally, we considered the importance of ``process over product'', which Schrock links to civic technology~\cite{schrock2018civic}, to maintain a focus on how civic technology represents not only an alternative means of developing technologies, but also a means for developing them for alternative purposes.

\subsection{Data Collection}
%Triangulation between observation/participation / interviews / document analysis

\begin{table*}
\begin{tabular}{|p{1.75cm}|p{1.75cm}|p{1.75cm}|p{1.75cm}|p{1.75cm}|p{1.75cm}|}
\hline
Interviewee & Active Organiser & Past Organiser & Project Participant & Active in another Project & Public Servant \\ \hline
1 & & X &    & X      &  \\ \hline
2 & & X &    &        & X \\ \hline
3 & X   & X &    & X      &  \\ \hline
4 & & X & X  &        & X \\ \hline
5 & X   & X &    & &  \\ \hline
6 & &   & X  & &  \\ \hline
7 & &   &    & X      &  \\ \hline
8 & X   & X & X  &        &  \\ \hline
9 & &   &    & X      &  \\ \hline
10    & X   &   &    & & X \\ \hline
11    & & X & X  & X      &  \\ \hline
\end{tabular}
\caption{Some of the roles played by interviewees at CTTG and outside}
\end{table*}

Data collection methods included a combination of action research (which included participant observation, attendance of meetings, and collaboration to run hacknights and other events), interviews, and document analysis. 

The first author conducted semi-structured eleven interviews with past and present members of the community and the organising team. These interviews, which lasted between 45 and 90 minutes, included organizers involved since the founding of CTTO, those who had stepped back from organising but were still active in the community, and organizers who had only recently joined into the community. The interviews focused on participants' understandings of the purpose and structure of the organization and their history of involvement, including discussions of decision-making, intention, values, and resources. In addition, special attention was paid to interactions between CTTO and other groups, especially with state actors such as public servants. Documents used in the analysis included CTTO blog posts and tweets from CTTO and project accounts, and documentation stored in CTTO's public shared drive. These documents included presentation slides from the weekly on-boarding presentation as well as documentation from organising workshops and meetings.

Action research is necessarily based on the involvement of the researcher within a real world situation. To this end the first author embedded themselves in the day-to-day work of organizing CTTO as a participant and organiser, as well as attending similar events and participating in city-run planning sessions. Prior to the official period of action research that commenced in Fall 2019, the first author already had extensive contact with the community, having attended hacknights since 2015 and participated in a few projects. Studying CTTO meant studying ``across'' the first author's peer group~\cite{clarke2005situational}, in a situation where the cost and risk to participation, and the power imbalances between researchers and participants, were minimal. During the main period of this study, between September 2019 and April 2021, the first author served as a co-organizer with CTTO, organizing and attending hacknights. Over the course of this time, that amounted to dozens of hours in organizing meetings, hundreds of hours attending hacknights, in addition to many months of discussions over Slack. As a member of the organizing team, the first author helped source venues and speakers, troubleshoot the pandemic-led transition to virtual hacknights, represented CTTO alongside their peers at conferences and on a podcast, and led several presentations and workshops at CTTO to present the state and focus of the research.

\subsection{Analysis}
\label{anal}
Methods for qualitative data analysis included interview and content coding, reflection, and techniques from situational analysis. Interviews were recorded, transcribed and coded as they were completed. Along with a mind to sensitizing concepts, interview codes highlighted participants views on the purpose and governance of the community, as well as their views on the relationship between the community and the state. Memoing~\cite{Charmaz2014} and journalling techniques that focused on interrogating the normative assumptions and boundaries of the research~\cite{Midgley2001} were used to ground the analysis throughout the data collection process. As potential themes emerged from reflection and analysis, interviews were coded again based on themes. Public-facing documentation from the community was also analysed in this way. In particular, discussions of how participation in the community benefited its members as well as the government, and how CTTO facilitated interactions with different civic issues and policy-making processes, emerged as the basis for this argument. As themes became more concrete in the form of publications, they were also discussed with the community at hacknights in the form of presentations and workshops, as well as individually with participants quoted from interviews.

\begin{figure}
    \centering
    \includegraphics[width=\textwidth]{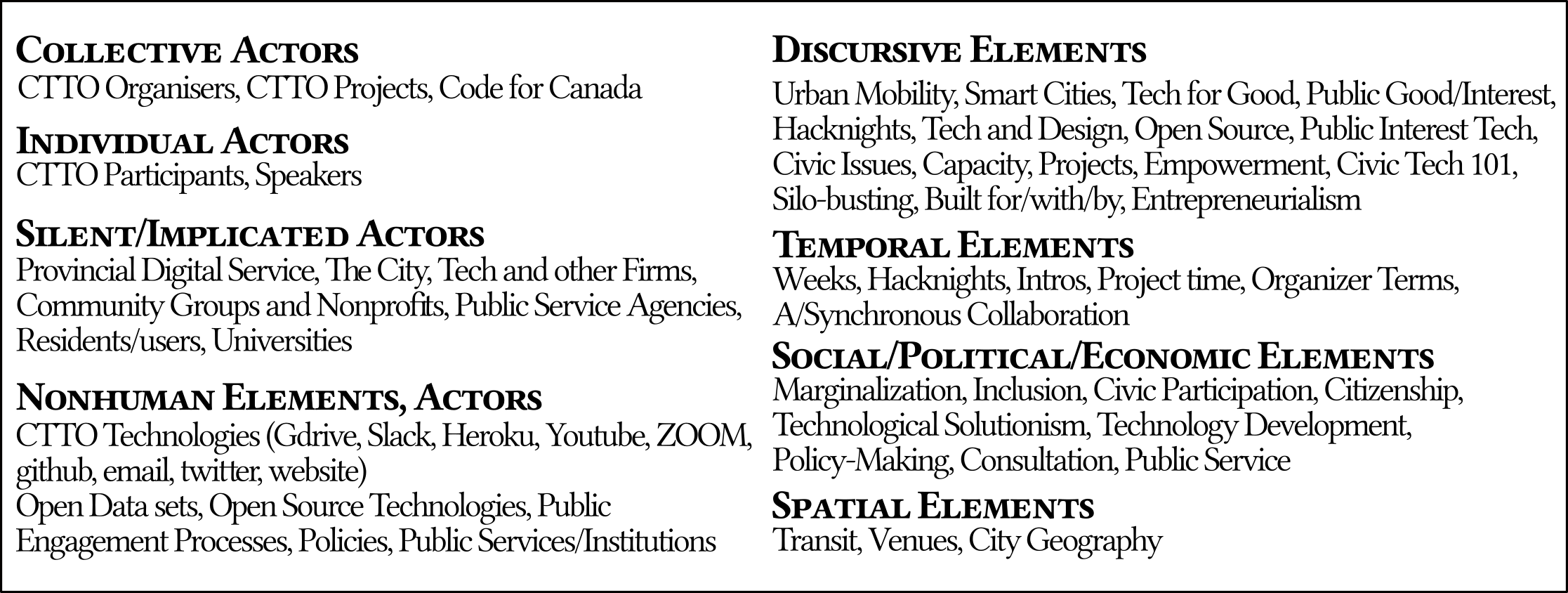}
    \caption{Ordered Situational Map}
    \label{sit-an}
\end{figure}

Situational analysis is a technique for analysing qualitative data through the iterated construction of hierarchical `maps' ~\cite{clarke2008social,clarke2005situational}. These maps were drawn throughout the process, in order to stabilize an understanding of the situation, and to structure reflection on research scope and emerging themes. The first kind of map, the `situational map', is used to stabilize a set of important elements in the situation and provoke analysis of their relationships~\cite{clarke2008social}. Fig.~\ref{sit-an} gives an ordered situational map that describes CTTO as a situation in terms of some of its elements. Using situational maps also helped to maintain the language of the community throughout analysis and reflection, putting community terms alongside, rather than subordinate to, academic ones. 

\begin{figure}
    \centering
    \includegraphics[width=\textwidth]{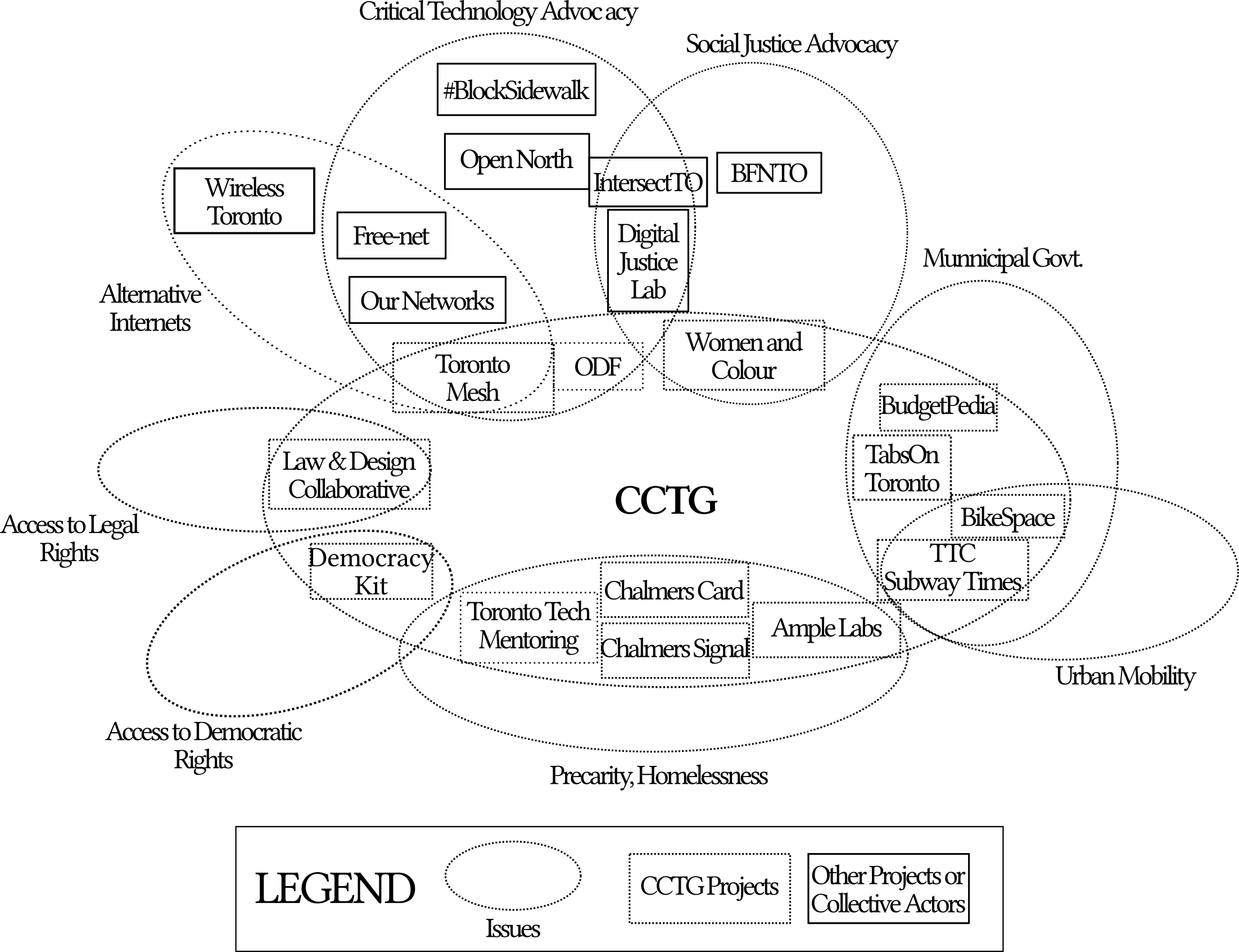}
    \Description[A social worlds/arenas mapping of CTTO projects to their issues, and including some other groups]{}
    \caption{A social worlds/arenas mapping of CTTO projects to their issues, and including some other groups}
    \label{fig:swa}
\end{figure}

The social worlds/arenas (SW/A) map shown in Figure~\ref{fig:swa} lays out collective actors in terms of their discursive and material commitments~\cite{clarke2008social}. To explore how projects originating at CTTG spoke to a variety of issues as their own kind of public, this analysis was supplemented by using a schema of issues and attachments adopted from work in CSCW, notably Le Dantec's work. Recall, an issue is the cause that a group is oriented toward, and attachments are the realization of that commitment in terms of the interventions and actions that a public takes toward its amelioration~\cite{le2016designing,asad2017creating,ledantec2011publics}. Mapping social worlds and arenas (collections of similar social worlds) around CTTO helped to situate CTTO within a broader set of collective actors focused on developing public interest technology. The SW/A map also documents the boundaries of our analysis, representing the scope of the study rather than absolute boundaries of the community. Additional studies would likely uncover other issues and document additional groups in existing arenas.

\subsection{Action Research and Claims to Knowledge Contribution}
% frame firmly within SA/GT/AR
The nature of action research, in particular the complicity that the researcher adopts in relation to community values and goals, means that it produces knowledge differently than traditional social or hard science research~\cite{checkland1998action,hayes2011relationship}. The first author, who conducted the fieldwork and analysis and whose reflections generated the themes leading to this paper, has been involved with CTTO since 2015, years before this research was conceived. This poses obvious difficulties for evaluations and critique of practises at the research site, but the aim of this research was never to evaluate CTTG’s practice in terms of effectiveness, impact, or other metrics. That would not be fair: it would hold the community to a standard that it does not aspire to. Even worse, it could be harmful to the community, suggesting that what these volunteers have built together is not enough. Rather, the research is intended to document the practises of the CTTO and represent an interpretation of that community to itself and to the academic world. 

In this manner, the research inevitably reflects the specifics of the relationship between the researcher and their peers in the community. As such, this research is not a total description of the community, it presents partial themes based on the involvement of a subset of the community's historical participants.  Other researchers could, even if adopting similar methods, interview more or different participants, and focus on different aspects of the situation. They may come to different interpretations and conclusions. However, we have tried to offer enough of a description of the method for data collection and analysis that the research can be considered `recoverable': i.e., that one can trace the argument back to the evidence it is based on~\cite{checkland1998action}. 

Action research is not aimed at making `generalizable' or absolute claims~\cite{hayes2011relationship}. This is not a comparative study where commonalities and contrasts can be leveraged to make claims general claims about context, important elements, etc. In this case, the basis for claims is constructed from a deep analysis of a single case, where interactions with participants have allowed us to represent the community for a scholarly audience. The methodology ought to be `transferable' to other, similar situations~\cite{hayes2011relationship}, as we have suggested above. Comparison can also occur through the relation of the work to other similar work within the discussion of the results, rather than in the findings. Comparative case studies could complement the present study in important ways in the future. 

\section{How does Civic Tech Toronto contribute to civic life?} \label{sec:CTTO}

\begin{figure}
    \centering
    \includegraphics[width=\textwidth]{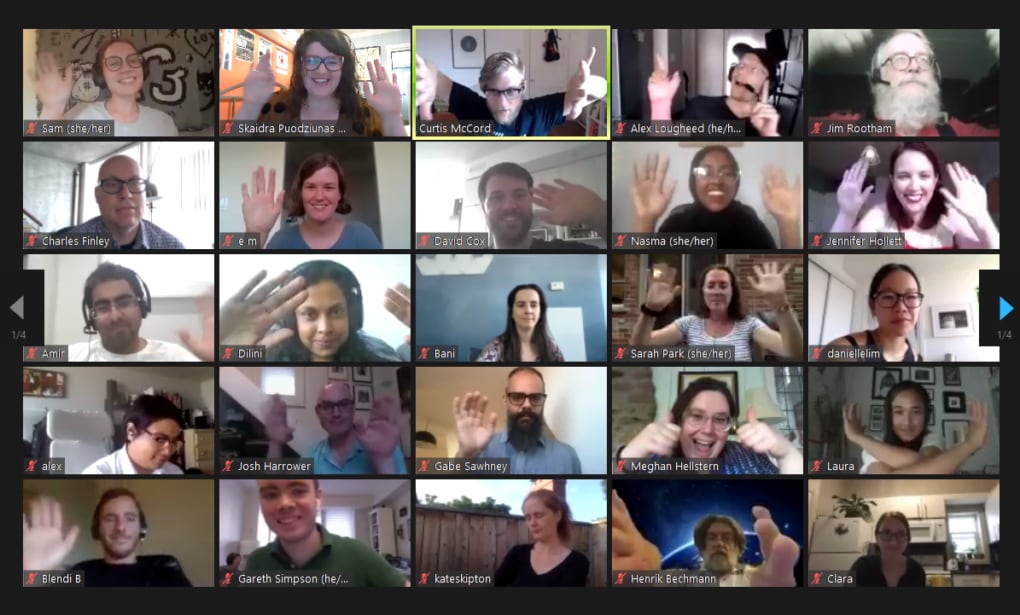}
    \caption{Participants pose for a photo during a virtual hacknight, hacknight \#250 photo serves as the current banner image for CTTO's website. Image Credit: Civic Tech Toronto (2020), Retrieved from civictech.ca}
    \Description[Screenshot from a CTTO virtual hacknight showing participants arranged in a grid. Participants are waving at the camera, posing for a group photo.]{Screenshot from a CTTO virtual hacknight showing participants}
    \label{fig:virt-ctto}
\end{figure}

In the summer of 2015, I (the first author) attended my first hacknight, a small gathering where people spoke knowledgeably and passionately about their city and its issues, and about how technology was, could, and would continue to change the way we lived together and what we could do together. 

Throughout the next five years, thousands of people would attend these hacknights, which became a way to explore spatial and social dimensions of the city.  We would ride elevators high up to corporate offices, climb flights of stairs into the studio offices of small technology companies, or step into enormous government halls. In these places, you could meet  designers, developers, artists, lawyers, master's and PhD students, community organizers, consultants, radio hosts, accountants, public servants from all three levels of government-- anyone really. All that united them was a curiosity and an interest in talking about their shared circumstance and how they could play a part in making their homes more liveable. 

The structure of the hacknights has not changed much over the years. They still begin with light socializing (although we cannot virtually share pizza) before proceeding to introductions. Everyone introduces themselves, no matter how many people there are. Next, a presentation followed by Q\&A. Following that, pitches, where members of the community get up to either make announcements about upcoming events or opportunities or to make a pitch for a project or idea that they want to discuss. They explain what the project is, its status, and what kind of skills are needed. For the first timers, organizers arrange a weekly introductory session that gives a sense of how the community works and what it is about. For the rest of the night people work together on those project or cycle around to different groups and conversations. When venues were sourced from among community members, these breakouts might be in conference rooms, workplace offices, or just whatever tables and chairs were available. Nowadays, breakout rooms suffice. At the end of in-person hacknights night the chairs are stacked and people head home or to a nearby bar to keep their conversations going into the night. Online, we all regroup in the main room for some friendly socializing. 

Civic tech groups operate in a number of Canadian cities, including Toronto, Montreal, Vancouver, Edmonton, Ottawa and Fredericton. The practise of civic tech varies among these groups. With only loose and occasional coordination between regional groups, local organizers draw influence from their peers in civic tech, from other social movements, from their members, and workplaces. Many are organized around events, called hacknights, inspired by Civic Tech Toronto (CTTO, which was itself influenced by the Chicago Hacknights \cite{chi2022chi}), the longest running civic tech group in Canada and the focus of our study. CTTO  has run hacknights steadily for over six years; it is run autonomously, without sponsorship, by volunteer members. Its membership composition includes public servants, tech workers, students, activists, and anyone curious about the group. By examining how projects at CTTO have sought to intervene in various civic issues, and how they have interacted with government and public bodies in the course of this work, we can begin to sketch a role for these types of communities in the work of democratic governance. 

By acting as a platform for collaborative civic projects, CTTO has supported a variety of participant-led interventions into a range of group-defined civic issues, which is how they understand the work of civic tech. CTTO has retained an attachment to civic technology that is centered on processes, rather than specific products. At the weekly hacknights, participants share food, listen to a speaker, and then join into breakout groups to work on projects. Some projects are specifically focused on intervening with public institutions and technologies. Some of these are more critical of governance processes. In other cases, projects are focused on addressing issues of marginalization, and others serve as examples of collaborations between the civic tech community and other organizations. As they interface with state actions or institutions, these projects demonstrate how a range of design and collaborative activities can be included within an expanded ambit of civic engagement. 

\subsection{Civic Tech acts as a contact zone for different social worlds}
% add ref RQ1
Contact zones are sites of exchange and collaboration between different social worlds, or ``interactive and improvised spaces of co-presence and encounter of previously separated social groups''~\cite[p.47]{prost2021contact}. Social worlds are defined by and differentiated from others through their shared sets of values, performances, and vocabularies. While communication inside that social world is adequately supported by shared meaning and implicit norms, communication between social worlds requires that the spaces and terms for this communication be set up. In part, these exchanges are linguistic, where participants (potentially) learn to understand the ways of speaking and terms of focus within other worlds~\cite{ratto2007contact}.  Learning and exchange is believed to result from productive tensions between the ways of thinking and working that different participants bring to a contact zone~\cite{prost2021contact}.

\begin{figure}
    \centering
    \includegraphics[width=.45\linewidth]{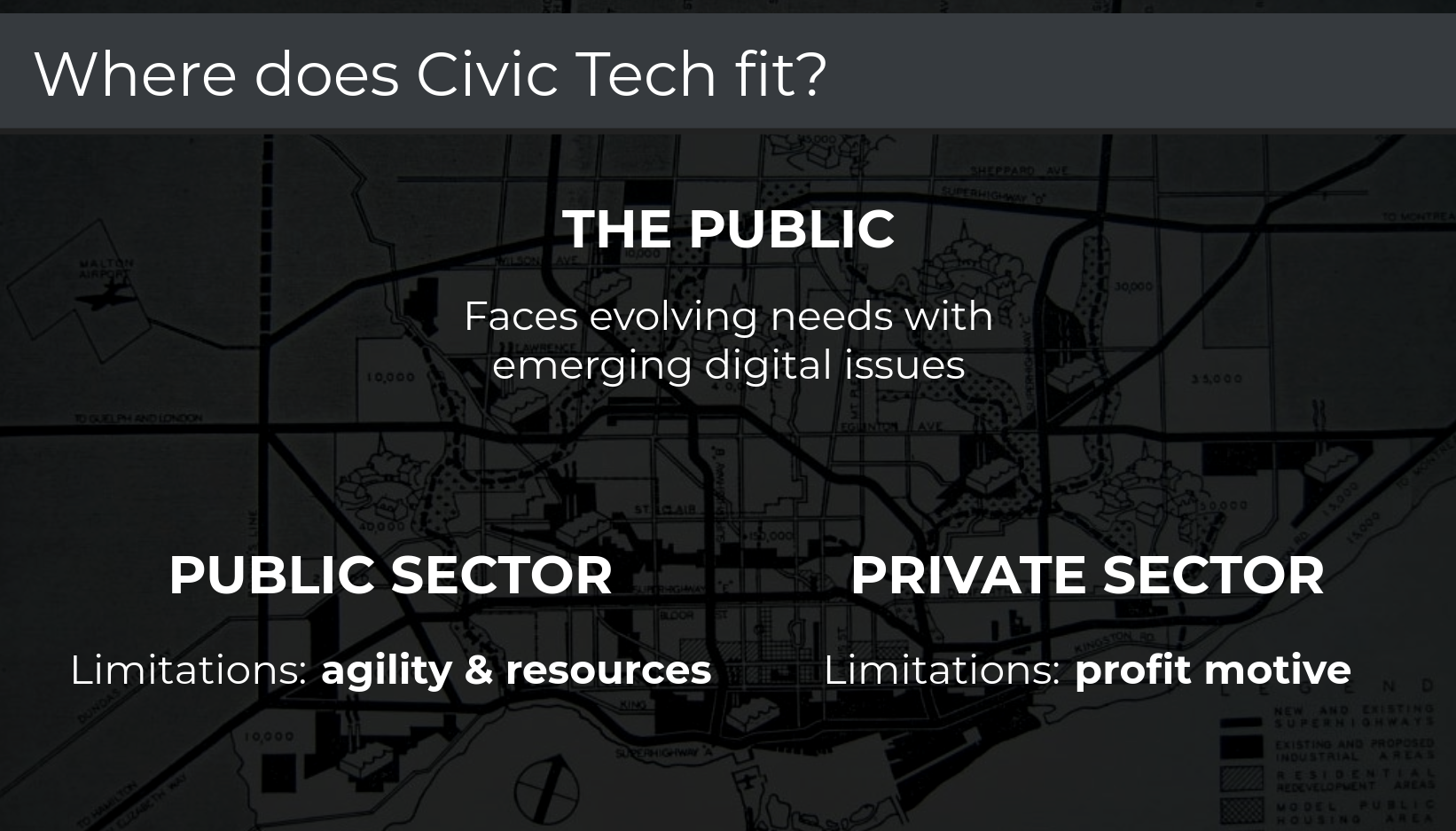}
    \includegraphics[width=.45\linewidth]{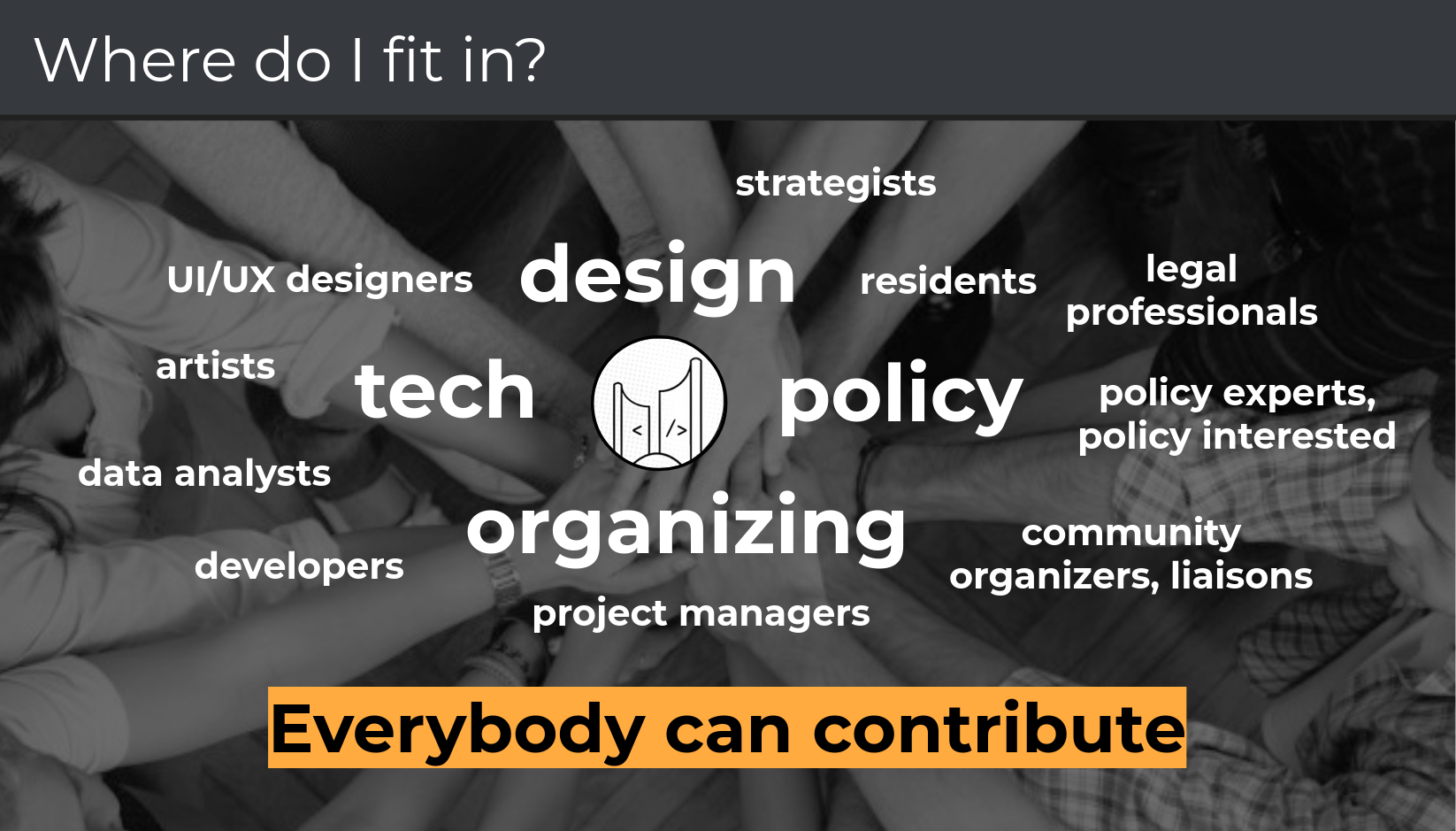}
        \caption{Two Slides from the onboarding presentation `Civic Tech 101', the left from 2018-05 and the right from 2021-04. CTTO explicitly positions itself as linking diverse expertise and domains to address issues of knowledge circulation and inclusion.}
    \Description[Onboarding slides]{Two slides from the CTTO onboarding presentation. The left slide is titled `Where do I fit in?' and lists roles like residents, designers, analytics, projects managers, community organizers above highlighted text reading `Everyone can contribute'. The right slide is titled `Where does Civic Tech Fit in' and  }
    \label{fig:ccto-fit}
\end{figure}

CTTO is designed to facilitate interactions between a diverse set of participants, especially between technologists, civil society advocates, and public servants. One founder, Gabe recalled hacknights, which began in 2015 amid a backdrop of new digitized approaches to urban mobility and the growth of open data, as a way of connecting technology expertise to the work of city-building. Gabe located some of the sluggishness they saw in the state's approach to technology as resultant from the fact that ``the lawyers don't know how to talk to the policy people, don't know how to talk to the programmers\ldots{} if we could create a space where these people with different skills and backgrounds and perspectives could come together \ldots{} then maybe we'll be further ahead'' (Gabe S).  

\begin{figure}
    \centering
    \includegraphics[width=.45\textwidth]{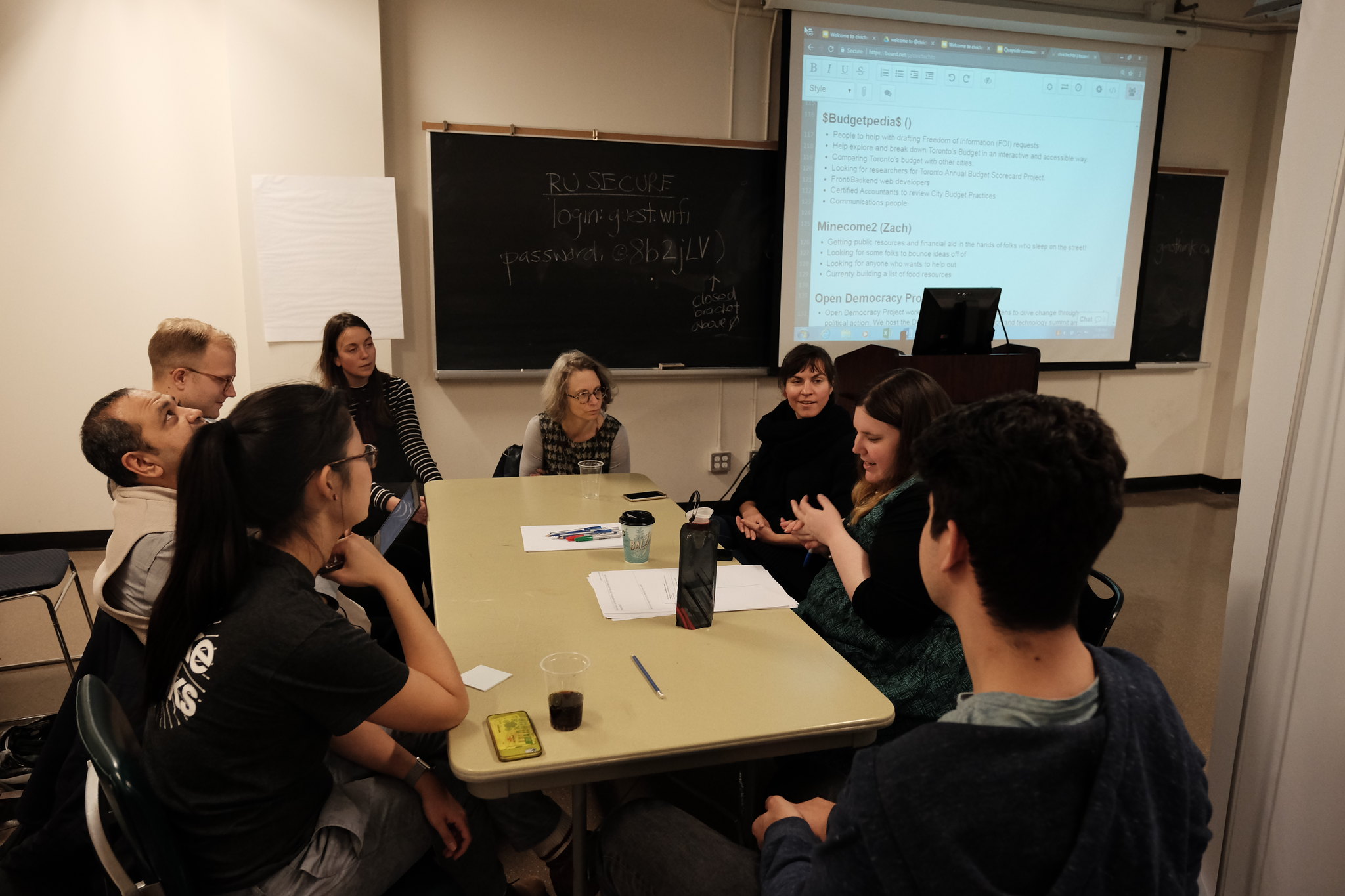}
    \includegraphics[width=.45\textwidth]{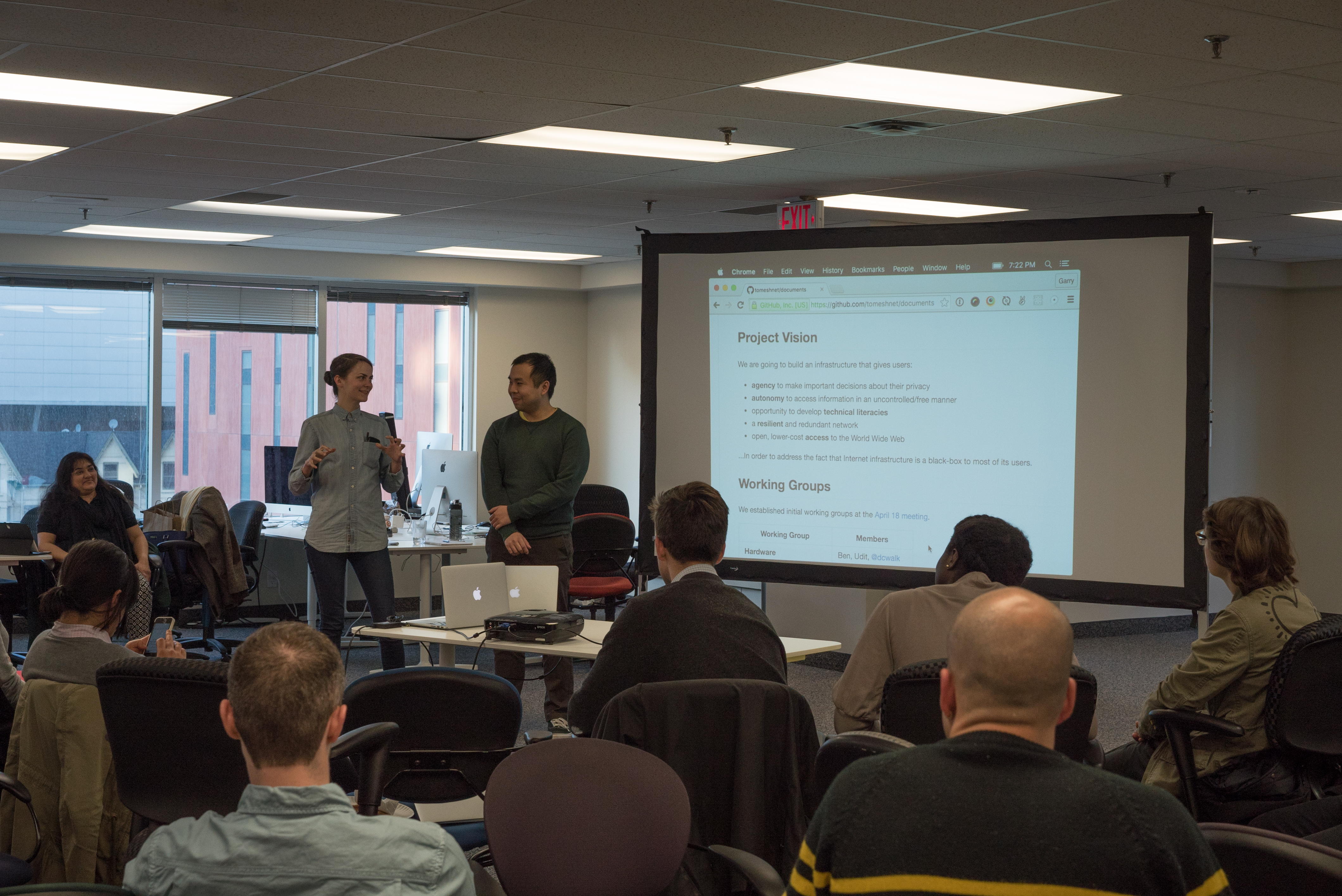}
    \caption{(Left) Hacknight attendees discuss a recently proposed Smart City project in 2017; at the table are academics, public servants, and technology professionals~\cite{ct-brainstorm2017}. (Right) Members from Toronto Mesh present the state of their work at a 2016 hacknight\cite{ctto201639}}. Credit for both images to Civic Tech Toronto.
    \label{fig:ct-quayside}
\end{figure}

One way by which CTTO moves between social worlds is through the rotation of hacknight venues. Prior to the virtualization of the hacknights, CTTO's venues were sourced on a month-by-month basis, with the only constraints being proximity to public transit and accessibility. Over its first five years, hacknights brought participants to a range of venues, including local technology companies, community spaces, the offices of major consultancies, numerous buildings at two of Toronto's downtown university campuses, and the very halls of government itself (See Figs.\ref{fig:ct-hacking} and~\ref{fig:ct-quayside}, below). This rotation creates a continual sense of novelty, allowing attendees to explore the sites where technology and governance work occurred, albeit after most employees have returned home. 

CTTO's weekly speaker series also helps to create an atmosphere of cross sectoral learning. While it is a fairly common opinion among organizers that speakers are not the main point of the hacknights, they remain important to structure the evening around a shared activity and to lower the barriers to entry. As another organizer, Skaidra, remarked in organizing discussions, speakers helped keep participation diverse and dynamic by bringing in their audiences and by helping to promote the event, which could get first time attendees to check out the community. This is well supported by experiences at the hacknights and by documentation on past speakers, including recordings from many of the hacknights\cite{ctto2022yt}, which features a wide range of people present, including software developers, academics, public servants, community organizers, other civic tech groups and projects, entrepreneurs, and artists.
%ctto2022yt

By acting as a `convener' in the words of one participant/organizer (Skaidra P), CTTO connects practitioners of various expertises, including tech and design, academia, civil service, nonprofits, and community organizing groups. What enables this exchange is a common civic framing, depoliticized but inviting discussion and action. It was exactly such framing that brought together over 100 people~\cite{ctto2018quayside} to discuss the the community's shared vision and plan of action around a recently proposed smart city project, or around the City of Toronto's emerging strategy towards digital infrastructure (discussed later in Sec.\ref{sec:prot-pub}). 

The positioning of CTTO at the boundary of different social worlds happens not only through the hosting of diverse speakers in diverse venues, but through the active reproduction of community values in one of CTTO's most important rituals, the weekly onboarding presentations known as `Civic Tech 101'. During the weekly onboarding presentation offered to first time attendees (Examples can be found in Fig.~\ref{fig:ccto-fit}), CTTO is positioned between three major spaces: the public sector, the private sector, and the public at large. In CTTO's terms, these social worlds are known as ``silos'', a metaphor that gestures to bountiful stores of knowledge kept institutionally separate and unmixed. Each of these silos is imagined to have its own strengths and weaknesses: the public sector is values-driven but often slow and ineffectual, the private sector is effective but distracted by a fixation on profit, and the public is faced with emerging (digital) issues and needs but has seemingly little power to address them. This is obviously a simplification, but it highlights the local ethos of civic technology: by complementing the perspectives and expertise of others, civic technology development strives to empower publics to address issues they identify in their cities. 

\subsection{Civic tech enacts alternative modes of technology development}
% RQ2  
\begin{figure}
    \centering
    \includegraphics[width=.45\textwidth]{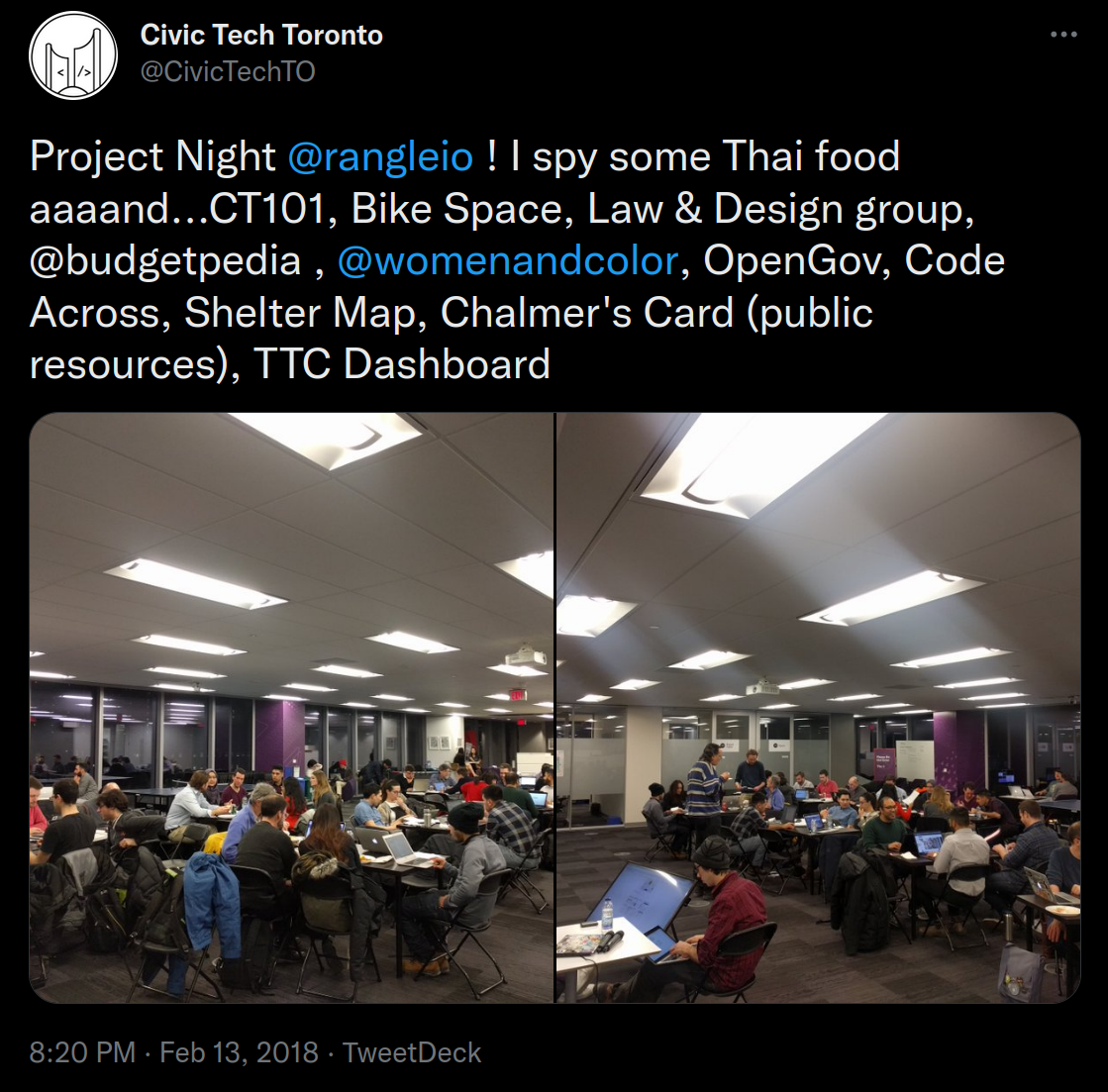}
    \includegraphics[width=.45\textwidth]{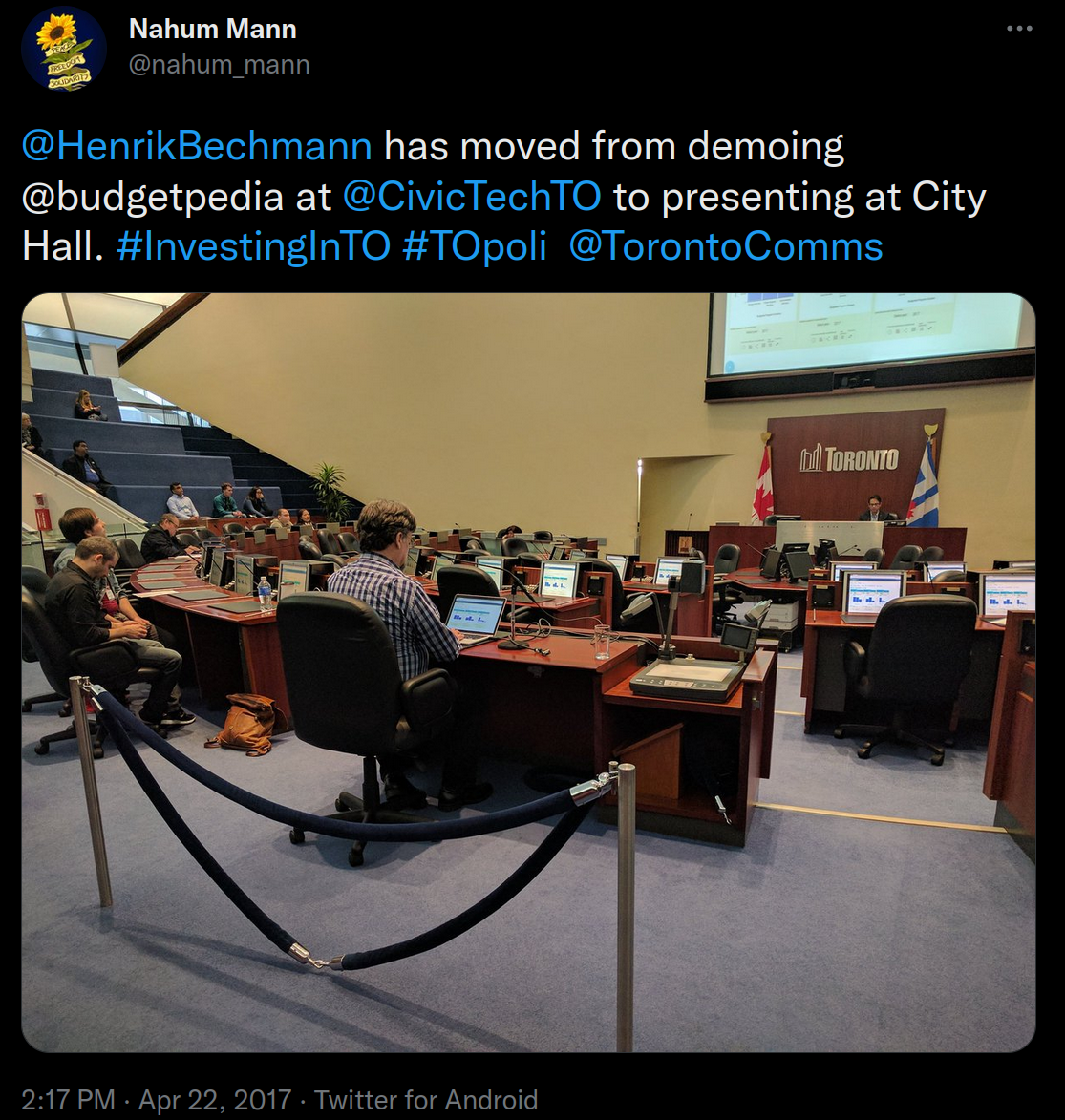}
    \caption{Tweets from CTTO's twitter and from a participant showing, typically bustling project time (Left), and hacknight presentations at City Hall (Right).}
    \label{fig:ct-hacking}
\end{figure}

% \usepackage{booktabs}
% \usepackage{multirow}
\begin{longtable}[c]{@{}>{\raggedright\arraybackslash}m{2cm}>{\raggedright\arraybackslash}m{4cm}>{\raggedright\arraybackslash}p{7cm} @{}}
\caption{A brief summary of \textbf{some} of the projects active during the first author's time at CTTO with the issue they address and the attachments chosen to address that issue}\label{proj-tab}\\
\centering\textbf{Project Name} & \centering\textbf{Issue} & \centering\textbf{Attachment} \tabularnewline
\toprule
\endhead
Amplified Mentoring & Precarity and access to opportunity & Brings technologists (especially from marginalized backgrounds) into contact with transitional housing providers to offer peer mentoring and support in technology skills for youth. \\ \midrule
Budgetpedia & Budget transparency & Aggregates and visualizes spending commitments and information to make the city's spending accessible and support improvements to the process. \\ \midrule
Toronto Mesh & Community wireless networking & A collective for experimental hardware and software development, also offering workshops supporting decentralized and community owned internet infrastructure\\ \midrule
Council-scraper & Access to Municipal Government & Scrapes city council transcripts to allow for keyword search and subscription notifications\\ \midrule
Women and Colour & Diversity and Inclusion in Tech & As a nonprofit, operates a social network and database of women and people of colour qualified to speak at tech events\\ \midrule
BikeSpace & Cycling infrastructure & Alongside Code for Canada and the City of Toronto, designed an application to report missing or damaged bike parking that would feed into the City of Toronto's work ticketing system \\ \midrule
Ample Labs & Access to essential services and support & Operated as a nonprofit providing (to government clients) a plain language chatbot and analytics platform that aggregates information about city services and can suggest services to people in need \\ \midrule
Law and Design Collaborative & Access to legal resources & Organizes volunteer experts from various fields to make legal information more accessible online. \\ \midrule
ODF & Data governance & Using interviews and storytelling, seeks to involves laypeople in speculative design about how data should be governed. \\ \midrule
TabsOn & Access to information & Supports keyword searching and email alerts of City Council proceedings using a web scraper.\\ \midrule  
Run4Office & Access to political rights & Provides information and resources to run for public office without major party support.\\ \midrule
Shelter Card & Access to essential services and support & Distributes durable, laser-cut cards containing information about basic services for those experiencing homelessness,\\ \midrule
Shelter Signal & Shelter capacity & Addresses major load-balancing issues in the city's shelter system by prototyping a free and open source hardware device to enable shelter staff to dynamically update shelter capacity on a public facing website. \\ \midrule
TTC Subway Times & Transit performance and data & Uses crowd-sourced metrics to evaluated the integrity of the municipal transit agency's estimated arrival time data \\ \midrule
\multirow{3}{*}{Code4Canada} & \parbox{4cm}{Technology capacity in \\ the Public Sector} & Fellowship program placing technologists within government projects with the aim of building in-house technology development capacity for smaller projects.\\ \cmidrule(l){3-3} 
&                                                           & User-testing service that compensates residents from marginalized backgrounds to help test government information systems and digital offerings.\\ \cmidrule(l){3-3} 
                               &                                                           & Offers a membership based venue for capacity building and workshops for public servants to increase technology skills.                                                            \\ \bottomrule
\end{longtable}

CTTO's major involvement with technology development is through the support of projects. Following the week's presentation, projects are `pitched' by participants, who describe the work they are doing and what kinds of contributions are sought. Generally, about half of the hacknights are devoted to project time; attendees spread out across the space, or into breakout rooms, where they talk and work until the end of the night (See Fig.~\ref{fig:ct-hacking} (Left)). Projects are initiated and maintained autonomously from the overall organizing functions of CTTO. Decisions about projects and their directions are not made or discussed at organizing meetings; project participants self-organize in hacknights and asynchronously, often through the CTTO Slack. As such, projects can take on a variety of characteristics and address a variety of issues. In its six years of hacknights, there have been a lot of projects hacked on at civic tech (see Table~\ref{proj-tab} for some examples). These projects range from technologies to access public information using chatbots, laser cut cars, and scrapers, to tech mentorship programs, supports to help people run for office, and experimental engagements in decentralized networking. 
% description of some projects here

Pitches are not always about fully fledged tech ideas; sometimes they explore concepts and facilitate discussions. Sometimes projects do not last more than a week. This is another feature that distinguishes the type of work that participants at CTTO undertake: there is no impetus to produce a product. The mode of technology development that animates hacknight projects are based on a degree of consensus between contributors, especially around the values that animate the articulation of the problem as a civic issue. A crucial contrast to the kind of technology development most familiar to many of the participants is that hacknight projects can be produced outside of market and bureaucratic constraints, such as funding sources, business models, deadlines, etc.~\cite{mccord2022civic}. Emily, an organizer and project contributor, was keen to point out that the lack of competition and pressure was important to make hacknights unlike waged work and to make space for creativity:

\begin{quote}
``One of the characteristics of civic tech that is important to me is that there's no money involved\ldots{} To me, it's really radical that we all get together once a week and there's\ldots not even the prospect of getting funded\ldots That's not how the private sector works where you're looking for a client to pay for something \ldots [or] how the public sector works, where \ldots{} we constantly have to be aware of the potential of audit and financial reporting and value for money\ldots I think it really creates a space for experimentation. And it also shows that things can get done without money.'' (Emily M)
\end{quote}

In some cases, projects do grow to a point where the work required to realize their ideas exceeds the time available at hacknights. Emily, themselves a participant in the BikeSpace project and organizer at CTTO, called this transition `graduation' and mentioned two other important projects, Ample Labs, and Women and Colour. As a project, BikeSpace emerged from a partnership between the City of Toronto and Code for Canada, enlisting volunteers from CTTO to develop a technology that would allow users of a web application to report places where bicycle parking (owned by the city) was either damaged or in short supply. This crowdsourced information could be made available directly to city staff. Volunteers worked at and between the hackights to produce this application, and with the help of a project manager contracted by the City of Toronto, produced a working application\cite{BikeSpace2020} that was even cloned by the City of Edmonton. Unfortunately, years after its release, questions about who would host the application failed to be resolved, and it has fallen into disuse. %BikeSpace2020
 
Ample Labs's main product was an AI-driven chatbot to facilitate information seeking and access to social services by processing plain language queries to search a database of municipal services and offerings. It also offered an analytics platform. Ample Labs's clients are municipalities who subscribe to the service, but its users include people experiencing homelessness or who are otherwise in need of information on social services, including first responders and professionals like social workers and librarians who can use this information to assist people in need. Ample Labs acquired considerable funding and support from partners that include major tech and telecommunications firm as well as public entities. In 2021, Ample Labs ceased operations, citing the difficulty in achieving organic growth of their use base and pressure from clients to seek more data from their users~\cite{chen2021note}.%chen2021note

In the case of Ample Labs, development had reached a stage where the technology was ready for market, and to offer it, the project teams incorporated as nonprofit entities. This allowed them to seek funding and offer services while retaining their focus on issue, but did expose them to the constraints of funders and clients. Ample Labs is not the only nonprofit that began as a CTTO project. Other examples include Women and Colour, Law and Design Collaborative (LDC), and Run4Office. 

CTTO's approach to technology development is based on the premise that a diversity of perspective provides a solid basis for the development of technologies intended to address social issues. To this end, participant engage in design activities removed from the exigencies of private and public sector technology production, while at the same time importing expertise from both fields, as well as from a variety of others (See Fig.\ref{fig:ccto-fit} (Right)). The presumption that a diversity of perspectives and expertise can lead to more thoughtful engagement with technology and its relation to civic issues has been baked into the practise of civic technology since it's early days. In a formative blog post, Christopher Whitaker defined civic technology not only in terms of the activities that the technologies were imagined to support, but as a new mode of development practise that involved participation from traditionally siloed areas~\cite{whitaker2015what}.

LDC, a collaboration between CTTO members and The Action Group for Access to Justice, exemplifies this approach. LDC also takes inspiration from CTTO in other ways, adopting a similar structure of organizing volunteer contributions to projects within a legal rights space, even applying the practice of combining complementary perspectives of civic technology to their own work, ``reversing the roles from traditional \textit{pro bono} legal work\ldots{}[by involving] professionals from\ldots{} user experience research and design, marketing, graphic design, coding, data science, consulting, education, law, and public policy\ldots{} to work on making incremental improvements to Steps to Justice, a plain language website that presents information about common legal problems in Ontario''~\cite{au2018civic}.

\subsection{Civic Tech creates a space for civic learning and experimentation}
%RQ1,2
As we have seen, the civic engagement envisioned at the hacknights is not just about creating a space for making contacts and discussion, but for making and design as well. As one of the founders, recalling their own view on the purpose of the group's inception, put it:

\begin{quote}
``[The] purpose was to create a sort of welcoming space where people with a range of skills, technical or not, and of knowledge, political or not, could come together to collaborate in a context that's very oriented around making. There are tonnes of events where people come together and talk, and we really wanted to have an event where people come together and make\ldots{} for the purpose of improving the city, or learning what we might need to know in order to have some sort of positive impact on the city'' (Gabe S).
\end{quote}

Some CTTO participants stress the importance of creating a space with a high tolerance for `failure', contrasting it with pressures for deliverables in the workplace. Projects cannot really fail within the world of CTTO, because products are not required to `succeed'. Organizer's exploring CTTO's purpose and theory of change in 2016 put it this way:

\begin{quote}
``Failed projects produce learning\dots [w]e’re not an engine for creating products and services -- we’re an engine for creating learning and community capacity. If great products and services come out of this community, that’s a wonderful plus, but we don’t want to live or die by the success of projects that emerge from our community''~\cite{Milito2016}.
\end{quote}

So, if the scope of a project becomes unmanageable after inquiry, or if the energy of the team dies down, there are no consequences to abandonment, and if work has been documented, it can be continued at a later date. Even if a project dissolves after some time, it still promotes CTTO's goals of creating relationships between participants around activities of making. Emily specifically connected CTTO's slower, more casual pace of experimentation to supporting learning among participants:

\begin{quote}
``For me civic tech is not about getting things done fast, because getting things done fast is not an inclusive way to do things\ldots{} [and] you're only ever going to be leaning on the people who are already really good at something and are already kind of experts\ldots{} [M]oving slowly makes space for people who aren't already experts, but also supports so much more careful consideration of how and why things are being done.'' (Emily M)
\end{quote} %  TODO this may be a good fit for the alternative modes section?

Hacknights also afford participants a venue to explore and practise technology skills, including web development, data analysis, writing, digital mapping, etc. Participants can also become acquainted with technology development processes more generally, gaining experiences in things like project scoping, team-based work, and version-control. Importantly, participants gain a deeper understanding of the situations and systems which their projects identify. In their time attending hacknights, the first author had numerous conversations with professionals and public servants who were looking for something more than what their everyday lives offered them, whether it was a use of their skills they perceived as more worthwhile, or just something they could choose. In the pursuit of their interests and causes, participants become accustomed to more egalitarian ways of working with each other, but they can also learn about the realities of what government works, and how that work is done. 

\begin{quote}
``I definitely felt that [the virtues of mass participation] applied to like, what civic hacknights were doing,\ldots{} people felt more engaged,\ldots{} [like a kind of] anti-apathy\ldots{} Civic tech spaces have that kind of free, unlocked leadership that's kind of flowing and moving, [and] everyone is learning. It's like they're learning about that thing that's changing, power, how power is a little bit like more fluid. And so you're learning a little bit more about how power works in the world. (Patrick C)
\end{quote}

CTTO also develops itself as a sort of technology. Hacknights, like mushrooms, are only the fruiting presence of a much more subtle network that grows below the surface. Governance at CTTO is fluid: roles and duties such as the organization of meetings are open to participants of any tenure. Because of this openness, the social reproduction of the community becomes a configurable opportunity for learning too. Taking on an organizing role is an opportunity for people to practise their public speaking, community organizing, social media, or writing skills. They can do so in a supportive environment, knowing that other members of the community are familiar with the roles and willing to step in and support them at any time. In their own time organizing with CTTO, the first author learned a lot about hosting online events, programmable Slack Bots, and presenting their research to broader audiences. 

Transactional products make activities legible to institutions, because they can provide abstracted and measurable evidence of success. While CTTO's process over product focus is important to creating the right type of environment for participants, it does mean that isolating the kind of transactive objects that make the work legibly productive to institutions (like universities or governments) is difficult. While CTTO has developed workflows to track speakers and participants, their difficulties lie in keeping up the bureaucratic capacity to manage these reporting tools. There are dozens of people who have given their time to organizing hacknights, hundreds of speakers who have presented, and thousands who have attended them. The meetup group boasts over five thousand members and the \#general Slack channel over twenty seven hundred. Regardless, participants at CTTO are enthusiastic in describing what they see as valuable about the community. Skaidra, a public servant, is clear that quantifiable metrics are not appropriate ways to evaluate the benefits of what the community has brought to government or civil society.

\begin{quote}
``I think the people who benefit from it are ultimately the people who leave with another idea, leave with something they hadn't considered before, and\ldots{} that's all you can really expect of this community\ldots{} I feel like people put a lot of pressure and expectations on volunteer leads\ldots{} especially when its civic tech, or when its govtech, to come up with a product, or show the value of the organization and\ldots{} I think [summarizing what the community has accomplished is done] through looking at who has come out to the events,\ldots{} how many conversations are being started,\ldots{} how many government officials are interacting with civilians\ldots{}[and] how big\ldots{} the network [is]\ldots{} [From the perspective of] digital government's community outreach, I think that's really what the benefit of CT is, being able to act quickly and mobilize community quickly and\ldots{}everyone benefits from that.'' (Skaidra P)
\end{quote}

CTTO hacknights provide low pressure learning opportunities that engage directly with the knowledges of the state. Through working on projects, attending presentations, and organizing work, participants can practise and learn a range of skills while contributing to the vivacity of the community. CTTO can also act as a site of exposure to some of the technological objects that are becoming more and more important to the work of governance and to the work of understanding and engaging with participatory processes. These objects include open data sets, online consultations, and government software systems as well as the more familiar policy documentation. 

\subsection{Civic Tech supports the emergence of publics and proto-publics}
\label{sec:prot-pub}
% RQ1
Ludwig et al. argue that as publics become more robust, constructing networks to cooperate and communicate with each other, they evolve into what the authours call ``communities\ldots{} [that] express themselves\ldots{} [and constitute a] general will''~\cite[p.194]{ludwig2016publics}. While it is somewhat difficult to ascribe a general will to a collective with a voluntary and consensual governance structure, CTTO appears sufficiently robust to meet that definition: it has a developed infrastructure to coordinate hacknights, a set of practises to reproduce the community over time, and a consistent focus on the intersection of technology and public life. So while Ludwig et al. argue that publics potentially grow into communities, CTTO provides an example of how a community can abet the emergence of publics. 

CTTO's social focus and structure is designed to connect people with similar interests and to encourage them to make inquiries, or take action, to address them. These groups share attachments to an issue through cooperative action, but not always focused on crossing the boundaries of their social world (civic technology) to create a broader coalition. They are more often ephemeral, existing sometimes for only a few hacknights, and are often focused on ideation and speculative design activities. In this case, it might be more appropriate to say that projects help constitute ``proto-publics'', which Lodato and Di Salvo identify as one outcome of the relationships formed at issue-oriented hackathons~\cite{lodato2016issue}. Even in the limited number of projects discussed already we find evidence of the emergence of proto-publics associated with transit and mobility, homelessness and precarity, access to rights, and digital infrastructure.

Some of these proto-publics do manage to grow into their own social worlds. Consider the example of Toronto Mesh. Founded in 2015 by CTTO participants (including the first author), this group set out to explore how alternative internet infrastructures could be used to challenge the existing centralized and privately controlled internet service provision in Canada. Activities included experimentation with existing technology, the design of new hardware/software combinations for networking (e.g. routers), and discussions of the value of decentralized infrastructures. Later on, Toronto Mesh moved away from the hacknights, as more time was needed for their work; they began running their own workshops, collaborating with the Toronto Public Library system and with makerspaces like HacklabTO, to introduce interested attendees to the concepts, software, and hardware of decentralized networking. In 2020, Toronto Mesh was in contact with employees from the City of Toronto regarding their public WiFi strategy (a need for which became more urgent as library spaces were closed during the pandemic). With additional contacts with a partnering networking firm, they began to set up `Supernodes' in parts of the city as part of their Community Wireless Network project. Nowadays, Toronto Mesh members routinely present information about their projects at hacknights~\cite{ccto2020tomesh} (see also Fig.~\ref{fig:ct-quayside}), while other members and their collaborators have created events such as the Our Networks conference for both local and global participants.

Through its role as a convener, CTTO also supports the emergence and activity of publics by acting as venue for public life and engagement with the state, for instance, by facilitating engagements with city consultations. In 2020 and 2021, several members of CTTO, including the first author, served on a Community Advisory Group for the City of Toronto's development of a Digital Infrastructure Strategic Framework, which also included discussions preceding the launch of a low-cost WiFi pilot program. During the public consultation phase, we convened workshops at hacknights following public meetings, inviting the civic tech community and other groups, such as members of the local ACORN chapter (a grassroots organization advocating for affordability in housing and internet access), to participate in a discussion of the plan and to share their priorities and concerns with the plan. At these workshops, we helped structure a reading of the materials and organized comments and discussions. Participants were also given information on how to submit their comments to council and how to sign up for a deputation at an upcoming council meeting. Later on, city staff working on development of the plan also gave a presentation at a hacknight and were available in breakout rooms for further discussions with participants. There, they were able to introduce attendees to the DIP and the technologies they were using to solicit comments from residents and took questions. While the results of this participation have yet to be seen, the hacknights were able to serve as a shared and neutral venue for contact between city staff seeking to communicate their work to publics more openly and engaged residents seeking to understand the work the city was doing and to share their values and perspectives with those employees. 

\subsection{Civic Tech connects people to the work of governance}
%RQ1,2
Through the use of technology and design skills, civic technology practise re-positions civic engagement outside of the discursive and formalized field of state run consultations and traditional advocacy, and it positions the artifacts and institutions of governance as a field for action. 

Civic tech projects are commonly based around interacting with open data using tools of query, display, and analysis to improve popular and expert understanding of policy and governance. While projects at CTTO do not always handle data in a way that appears cutting edge, it nonetheless uses and creates open data as a source of innovation. For example, Budgetpedia aimed to use data visualization technology to understand the Toronto's budget and spending decisions. However, because the City's budget is not given line by line in any consolidated set of documentation, project members spent years understanding how to collect and combine information from the budget into a coherent visualization schema. This also meant creating documentation around the budget process and concepts to ensure that readers could engage with this (ostensibly) public information from a variety of starting points. This transparency work can also be critical; one of Budgetpedia's main findings from investigation of the budget was that over time, spending decisions demonstrated a preference for capital funding projects, while the relative share of the budget devoted to operating expenses was shrinking year over year. This kind of observation was not possible outside of the backdrop of Budgetpedia's aggregation and parsing of municipal data and documentation. 

Another example of creating a continuous interface between civic technology work and governance is through a transit dashboarding project\cite{ctto2022ttc}. Meeting on and off for years, TTC Subway Times has constructed a system that compares crowd-sourced data on subway arrival times with the Toronto Transit Commission (TTC)'s estimated timing to help validate the performance of the transit system. A longtime steward for this project is also a public servant, working as a data analyst for the the City of Toronto while this project was running. Participants interested in transit were not only able to benefit from his technical expertise in mapping and data analysis but also from his access and knowledge of transit data more generally. 
%{ctto2022ttc}

This kind of interaction with public servants is an important development abetted by civic technology groups like CTTO. As one interviewee observed, this stands in contrast with other civic tech groups whose main point of contact with the state is through elected representatives. 

\begin{quote}
``I wasn't imagining myself working with [public servants]\ldots{} Even when visiting other kinds of sister groups, running their own hacknights, I feel sometimes there's still a kind of undercurrent of frustration at the public service. I think when they're not represented in the community itself, they're seen as outsiders\ldots{} [In other civic tech communities,] the people are involved in campaigning, electoral processes, the people who attend tend to be very engaged in those processes\ldots{} It makes sense: the electoral process, the campaigning process, is very permeable compared to the the normal public service. And so people see that as the site of action, and they have a lot of empathy for the politicians. They're supporting the politician, whereas in Toronto \ldots{} because the public servants were involved early, [the public service is] seen as a site of action more.'' (Patrick C)
\end{quote}

Interactions with public servants, as opposed to representatives, bring civic tech groups closer to policy-making processes that are becoming increasingly important, like the design of software systems, the release and promotion of open data, and engagement processes leading to the drafting of legislation. As participants gain important literacies, they become more effective collaborators with public services, both within the informal space of CTTO as well as within more formal interactions. The activities and relationships that emerge from CTTO can also been seen as useful to public servants, legitimating and even strengthening their work. Speaking in their capacity as a public servant, Skaidra put it as follows:

\begin{quote}
``Its one thing to do the work of bringing simpler, faster, better services\ldots{} but it's another thing to actually build ambassadors\ldots{} CTTO volunteers are kind of like the reason that digital government exists in some ways, you know? They are often the community who understands what government is trying to do, they're often members that are trying to support government when they are doing things a bit differently,\ldots{}[that] adds a layer of legitimacy to the work that we do, because people aren't getting paid to `endorse' the work that we're doing, but they're doing it because they know its right and because they see value in it\ldots{} [T]he community partners with us on things, and I mean we haven't done a lot of that officially, but I would like to say that unofficially we partner all the time, you know, just the fact that we have Ontario Digital Service members that speak at nights, to talk about our Open Data Catalogue\ldots{} go out to these nights\ldots{} talk about them\ldots{} Share it with their colleagues \ldots{} and invest their personal time, and professional time.'' (Skaidra P)
\end{quote}

Cultivating relationships with civic technology groups also serves the interests of the state by bringing technology expertise into closer proximity to the public sector. Governments at all levels throughout Canada are now actively trying to increase their capacity to build technology in-house. This has meant expanding the types of expertise in the public service, complementing the role of IT departments, and demanding new hires and major developments in the organization's capacities. 

The social world of civic technology has played a fore-running role in building these capacities in government. Code for Canada, a nonprofit group formed by CTTO organizers in 2016, the same year as the Ontario Digital Service, has run an annual fellowship program devoted to placing technologists into government projects on a temporary basis. Each year, about a dozen technologists are embedded into government teams, bringing with them sought-after expertise. In Fall 2021, the provincial government launched its own fellowship program and presented it at a CTTO hacknight. CTTO is one potential pool to draw applicants from. For years, CTTO participants have transitioned into public sector employment as they learn more about government and make connections at hacknights. Former and present CTTO participants and organizers are now present in all three levels of government, municipal, provincial and federal. 

\subsection{Summary and Reflections}
This section has shown several ways in which Civic Tech can contribute to civic life (RQ1): by acting as a contact zone for different social worlds, by enabling new experimental spaces, and connecting existing actors in new ways. Exploring the case of CTTO also showed how Civic Tech can reconfigure the roles and relationships of participating actors (RQ2): by enacting alternative modes of technology development, creating a space for civic learning, and creating new connections with public servants. We explored how hacknights have acted as a stable venue for residents (and now, virtually, anyone with internet access on a Tuesday night) to perform their political subjectivity in a way that is not generally afforded by the institutions of representative democracies. CTTO thus posits a kind of political subjectivity that is based in community and social interaction, and one that locates civic action in engagement with design, thereby expanding the set of activities we can understand as civic and carving out new roles for participants and residents in the processes of democratic governance. As a weekly feature in Toronto for over six years, CTTO hacknights enact a view of civic engagement that is based on cooperation with fellow civilians and with public servants. Participation in CTTO is casual, social, nonpartisan, experimental, and flexible:

\begin{enumerate}
\item \textit{casual}: attendance is not compulsory, and organizing duties are structured to be manageable for volunteers;
\item \textit{social}: participation is based on finding mutual interests with peers, with socializing time before, during, and after the events;
\item \textit{nonpartisan}: the group itself is unaffiliated with any political party -- civilians interact with public servants who work for the government of the day and are ethically restricted from professing partisan affiliations;
\item \textit{experimental}: activities are based on imagining and designing technologies with no requirements for delivery; 
\item \textit{flexible}: CTTO is run by volunteers, without static venue or funding sources.
\end{enumerate}

Some of these qualities were very helpful in managing the rapid transition to virtual hacknights in March 2020. Organizers were quick to identify the need to cease in-person congregations and eventually selected Zoom as a software for facilitating virtual hacknights (as it had quickly become the software of choice for such matters). Thanks to continual experimentation with different technologies for captioning, breakout rooms (before Zoom allowed participant to self-select breakout rooms) and collaboration, hacknights have continued virtually for now over two years. To an extent, the existing rituals for making the community inclusive, like time for unstructured socializing, introductions, and Civic Tech 101, translated reasonably well to an online setting and continued to support new participants. The same can be said about technologies used to facilitate asynchronous collaboration in pre-pandemic hacknights. In this time, CTTO has made changes to its hacknight technologies and to the roles that organisers play in the hacknights. For example, hacknights now include multiple chat moderators, who post links into the chat and can answer questions, and Emcees, who host the hacknights, now use the chat pane to queue up questions and speakers.  Collaborative document software and messaging systems were already a part of the reproductive labour of CTTO and continued to facilitate the community in organizing virtual meetings.

That being said, this transition to virtual hacknights has not been without friction. For one, virtualization of hacknights using software commonly deployed in workplaces was perceived by some as making the hacknights much more like a workplace, and there was inertia as people adapted to the drastically different experience of video calling (for example, through so called `Zoom-fatigue'). In planning conversations, organisers identified the need to reinforce the social nature of the hacknights, and some lamented the difference in feelings. Virtual hacknights have struggled to simulate the vivacity and fun of in person hacknights, although speakers and projects continue to bring in new and recurrent participants. The change is somewhat tangible in the projects: new ideas and the energy to begin them are sparser. Although there are notable example of technology projects that have seen continuity and even reprisal, breakout groups focusing on discussion were more common for a time. Ultimately, enthusiasm about sharing and collaboration is essential to keep projects interesting and to turn ideas into projects, but these require high energy that was difficult to muster throughout two years of ongoing pandemic conditions.

By emphasizing a community driven model that is casual, social, and focused on learning and doing, the hacknights place civic engagement within the activities of everyday life. These instances of contact and collaboration help to build civic literacies in the public, while also giving experiences to public servants in presenting to, and working with, civilians. 

Focusing on the roles, activities and processes that count as civic engagement helps to demonstrate some of the major contemporary obstacles to designing (actually) participatory processes. Notably, individuals often become enrolled in pseudo-participatory institutional engagement processes that are transactional, soliciting information from them without increasing political agency or efficacy. Some ways of overcoming the limitations of a transactional focus can be found by focusing on how publics exercise agency in design processes and maintenance, working with each other, across infrastructures.

\section{Discussion} \label{sec:discuss}
\subsection{Civics are Sociotechnical: Beyond Transactional Democracy}
What is meant by calling forms of civic engagement `transactional'? Researchers in the field of Digital Civics identify transactional logic in play in governance approaches that are focussed on a data-driven approach to delivering `services' which position residents as `users', consumers, or even `customers' of public services, products which the state provides~\cite{cardullo2019being, dickinson2019cavalry, asad2017creating, vlachokyriakos2016digital, mcdonald2019}. Equating governance with the planning and delivery of services construes political subjects as relatively passive, but as Asad et al. astutely point out, this transactional logic also underlies seemingly participatory processes based on ``rational and deliberative frameworks''~\cite{asad2017creating}[p.2034]. To understand and overcome this transactional logic, it is necessary not only to ask what activities are afforded, and what arrangements of power are supported, but also what roles and relationships they enact for political subjects. What kinds of democracy are supposed by the technologies our states employ? What kinds of imaginaries are performed, what roles given to people? Without committing to a philosophical analysis of the transactional logic within deliberative democratic theory, we also use the term `transactional' to describe how interactions between citizens and states involve the elicitation of discrete pieces of information that can be handled by institutions. 

When is community engagement transactional, rather than relational? As discussed above in~\ref{sec:background}, the answer lies in understanding how technologies and processes configure people's agency and power in relation to the outcome of an engagement process, as well as to the process itself. Palacin et. al use to concept of `pseudo-participation' to provide a more precise articulation~\cite{palacin2020design}. Technologies and processes are value-laden, and strictly determine the roles that participants may play in an engagement process through sets of technological affordances (this includes, for example, agendas for meetings, question times, breakout group topics etc.). But processes themselves also structure participation in ways that the role of participants is marginal, known as pseudo-participation in design~\cite{palacin2020design}. There are many different kinds of civic acts that take this form. Contributing to a crowdsourcing project is one such example, whereby states elicit, using a specialized information systems, a structured contribution from residents or citizens, and then use this information in their decision-making. Voting also takes this form, whereby discrete points of information are elicited, and based on the reckoning of the result, a specific action occurs. Such processes do not provide agency to participants, because the role of people is to provide information for a larger process to function. As McDonald et al. write, these processes ``limit the possibility for citizens to be recognized as legitimate actors within the policy-making process''~\cite[17]{mcdonald2019}. These sorts of engagements, like consultations, individuate people, asking them to speak for themselves, based on their immediate opinions, rather than to learn about complex issues through interaction with public servants and their communities. In transactional processes that marginalize people in this way, decisions to act are (ideally) mechanistic/deterministic, causally resulting from inputs, and providing those inputs is the extent of agency afforded to citizens, a form of pseudo-participation in design~\cite{palacin2020design} or tokenism~\cite{cardullo2019being}. In some cases, this type of relationship may be appropriate to the task at hand, but these kinds of engagements cannot be said to be meaningfully involving people in democratic processes.

%TODO EXPAND
The transactional and pseudo-participatory form also underwrites the logic of processes that appear more complex, and even collaborative. When participants lack the ability to question the process and are immanent to an institutional agenda, this can be called pseudo-participation~\cite{palacin2020design}. For example, even in the case of Alphabet Sidewalk Lab's extensive public engagement programme in Toronto, the boundaries of the proposal were carefully constrained by decision-maker. While resident input and collaboration could yield changes to minor, occasionally even significant elements of the plan, control over the overall vision and purpose of the plan were ultimately maintained by Sidewalk Labs' authority~\cite{McCord2019}. From the perspective of participants, these kinds of experiences can be alienating and erode trust in participatory processes~\cite{corbett2018exploring}. 

Following emerging research on digital civics, we are interested in taking a `relational' approach to the development of state services, systems, and policy, where civilians can exert more agency in state-led processes of technology design and policy development~\cite{vlachokyriakos2016digital,corbett2018exploring}. Realizing goals of stronger participatory democracy, according to scholars like Dickinson et al., and Asad et al., requires a different approach to (non-transactional) civic technology, one that is focused on cultivating relationships between community members and groups and between those groups and their local governments~\cite{dickinson2019cavalry,asad2017creating}. Dickinson et al. show how technologies of asset mapping can be useful in building community power that increases the effectiveness of intra-community action and builds capacity that communities, especially those historically marginalized and under-served, can use to interact with the state. The findings from research with CTTO show how another technology practise, the hacknight, can also move toward a more relational approach to governance through civic technology. 

\subsection{Infrastructuring Relational Civic Technology}
CTTO builds relationships by providing a venue for civic engagement that is not constrained by the need to produce information for institutional use. The kind of relations that groups like CTTO associate with civic engagement and participation are grounded in the community performance and personal fulfillment through projects where participants can contribute their experiences and expertise, but without obligation to deliver. CTTO organizes a place and a time for groups as well as individuals to imagine and enact interventions into the issues they observe in their everyday life, and also brings them into closer proximity with the work of local governments and other civic tech organizations around the world. To do this, CTTO has a built an infrastructure out of sets of technologies and social rituals that have maintained a large community over several years. CTTO and other groups have a range of rituals that are used to reproduce the community and its ethos; every hacknight includes reference to the code of conduct, introductions by all attendees, a session where new members can learn more about how the group works, and social time after the hacknight. These rituals situate participants alongside their peers, who are there to share their knowledge, perspectives, and time freely. Project work requires participants to come to a shared understanding of their chosen problematic situation and work to imagine an intervention. This gives participants a degree of agency that is not typically afforded in processes that are facilitated by consultants or public servants on behalf of external decision-makers. Even in the cases where CTTO organizes participation in existing state led projects, like consultations, participants can engage with these processes from a community base, and supports them in understanding, and critically engaging with these processes~\cite{peer2019workshops}. 

\subsubsection{Hacknights, Trust Work, and Partnerships}
Building a relational civics requires that trust be built between publics and public servants~\cite{corbett2018exploring,dickinson2019cavalry}, or what Corbett and Le Dantec call ``trust work''~\cite{corbett2018going}. Exploring how trust can be built over time by interviews with public servants, Corbett and Le Dantec delineate a number of activities that build and retain the types of relationships needed to support a relational civics~\cite{corbett2018problem}. Civic engagement, within a city or a province, is not simply a discrete set of processes. Rather, these instances of contact and engagement are historical processes over which trust can be gained or lost over time~\cite{corbett2018exploring}. Creating and maintaining stable relationships with these communities supports the role of public servants as facilitators of participatory processes ~\cite{weise2020infrastructuring}, by building trust, creating more avenues for feedback and discussion, and helping to amplify ongoing work. While the term civic technology inevitably draws attention to specific kinds of artifacts, the example of CTTO and other civic technology groups in Canada points to a crucial inversion, that the civic itself is created and maintained both as, and by, technologies. CTTO and other groups like it represent sociotechnical configurations of civic life, positing sets of activities and relationships that can build trust and capacity for future civic engagement. 

To begin, hacknights act as an informal venue for public servants to present their work and receive feedback in a casual setting where participants are already familiar, a way of ``raising awareness...[defined as] making municipal operations legible to city residents... making sure that residents are aware that a department exists as well as its function''~\cite{corbett2018problem}[p.574]. Presentations by public servants help to demystify and make state business more visible (a form of civic learning, discussed again below), but they also make the people working on those projects visible as well, ``building relationships...[through] direct, personal contact between municipal officials and the public to whom they are accountable''~\cite{corbett2018problem}[p.574]. These points of contact can allow public servants to make shows of good faith, and share resources with potential participants in later formal engagements (``setting the table'' and ``finding opportunity''~\cite{corbett2018problem}). They may even be a good opportunity for civilians to influence the design of participatory processes, which Weise and Chiasson argue is an important step to realizing greater agency within these processes~\cite{weise2020infrastructuring}.  The relationships built at hacknights carry over to official processes for public engagements, where public servants and participants are known to each other, perhaps even with a history of collaboration or discussion. In a sense, this kind of relationship is more honest, allowing public servants to communicate their mandates and agendas, what Corbett and Le Dantec call ``mediating expectations''~\cite{corbett2018exploring}. So, while participation in hacknights is not civic engagement in the sense of participation within state sanctioned systems, it nonetheless bring participants closer to these systems, equipping them with knowledge and collaborators that strengthen their ability to act. Hacknights demarcate a stable time and place for the performance of public life that is located in social and individual fulfillment, and in this way we can call it a platform for civic engagement.

While groups like CTTO are recognized \textit{as audiences} for the government, they are harder to incorporate \textit{as collaborators}. They occupy an intermediate space between engaging individuals and contracting with firms for services. Recognizing these civic tech communities as a valuable interlocutor for states and governments should encourage states to support them in some way, while preserving their autonomy. Mustering and maintaining communities like CTTO takes labour and resources. While the voluntary nature of CTTO is at the core of its ethos, supports could help the community grow and maintain itself. Within CTTO's own history, support in the form of expertise (in project management, communications, booking and partnerships) as well as in terms of resources, have had positive impacts on the community. Occasionally, this expertise comes from the public servants who volunteer for the community, or from Code for Canada, whose employees and fellows have served a variety of roles in CTTO's organizing group. In part, what grassroots civic tech groups in Canada often lack is a guarantor, something to assure their success or longevity over time. 

Historically, CTTO's success is assured by a steady pool of volunteer organizers and in-kind support from venues. Realizing the potential benefits of this networks of learning and collaboration over time depends on the work of infrastructuring: it makes them durable and maintainable so that they can serve as sites for action in the future ~\cite{le2016designing}. All involved actors must learn how to do the infrastructuring work of building relationships between government and civic groups~\cite{crabu2018bottom}, continually rehearsing and maintaining those connections. In the case of CTTO, there are a few opportunities for support; promotion, which can be done by governments, universities, and other groups, technical support, such as the provision of code hosting, contribution of expertise, and offering up venues and food sponsorship. Association with larger organizations can make civic tech groups more stable over time. For example, Civic Tech Fredericton benefits from its proximity to entities like Greater Fredericton Social Innovation~\cite{mackinnon2021civic}. Other groups, such as Medialab Prado, exemplify how the benefits of stable funding, staffing, and even venues can have a major impact on the capacities of civic tech groups to organize events, find speakers, and coordinate projects~\cite{garcia2021dci}. Code for DC has built several projects in cooperation with nonprofit groups, owing to a high degree of skill in partnership management on their organizing team~\cite{c4dc2020civic}.
% c4dc2020civic

While such guarantees and supports could increase the capacity and influence of groups like CTTO, there is a danger that to become more legible to their supporting entities, they will need to sacrifice some of their autonomy in order to produce more recognizable, transactional outcomes that can be reported. As Holdgaard and Yndigegn point out, navigating the different priorities and values of cross-sectoral partnerships can trouble participatory design processes, especially when participation begins after an initial (and deciding) vision or mandate has been produced~\cite{holdgaard2018participatory}. This has certainly been the case in discussions around the Digital Infrastructure Strategic Framework, where differences in what public servants received as a mandate from elected representatives frequently did not align with community needs. A similar tension between initial vision and participation also afflicted the recent Sidewalk Labs development in Toronto, Canada~\cite{McCord2019}. 

Partnerships with communities like CTTO are also put under strain later in technology development lifecyles, especially in differences in capacity and attachment intersect with issues like maitenance. Keysar et al., for example, discuss some of the difficulties in using a participatory process to design community networks in Berlin. Not only were there major obstacles at the beginning of the project, but the nature of the funding and project commitments by state and university actors left major questions about who would bear the stewardship and maintenance costs after these larger entities had completed their involvement~\cite{keysar2022prototypes}. BikeSpace was met by a similar fate in 2021 with questions about who would maintain and host the application code. Many of the volunteers from the project having since moved on, and given that CTTO has no permanent material means or technical capacity to accomplish this, and because other partners did not find their own reasons to steward the project, the application is now inactive.

Without a robust internal decision-making process that involves all its members, it is likely that CTTO would struggle to effectively engage with larger and more powerful entities like states and firms to produce technology products. As it stands, CTTO is autonomous because its decision-makers, organizers from the community, have no reporting or production mandates. This is only possible because CTTO does not have a legal status. While incorporating as a nonprofit would enable the group to secure more funding and perhaps even engage in partnerships, this is not a direction that CTTO organizers have seriously pursued. 

%added holdgaar and keysar cites
% TODO update DIP name???

For some at CTTO, the eschewal of outcomes and measures makes CTTO hacknights a subversive space. One participant's description heavily alludes to Lindtner et al.'s~\cite{Lindtner2018} concept of ``parasitic resistance,'' where alternatives are built from within existing power structure be appropriating resources to their own ends, without the exploitation of labour that hierarchical production relies on:

\begin{quote}
``Civic tech is about gaming the system in the best possible way and playing within the temporary structures set up by capitalism to show that something else is possible. You know, there are super nice office buildings that have spare cash,\ldots{} [and] I have no problem with redistributing some of that abundance towards a process where I think the decision-making is much more shared, than it would be in say, a boardroom. \ldots{}It gives a lot of people a sense of purpose, that they may not get from their day jobs.'' (Emily M)
\end{quote}

\subsubsection{Process over Product}
Moving from transactional to relational models of civic engagement and civic technology requires de-centering the technological artifacts produced, and instead focus on the activities that occur around these technologies. This echoes Whitney et al.'s criteria for a social justice oriented and critical HCI that can adequately address the political, something they believe can only be done by eschewing the ``technological fix'' in favour of analysing how activities of design can create stronger social relationships~\cite{whitney2021hci}. 

In recent years, another popular site for the cooperative development of civic technology has been the hackathon~\cite{johnson2014civic}. Hackathons, in brief, are technology sprints, where a host convenes a set of programmers and subject matter experts to develop a technology product in response to a prompt, often over a weekend. Hackathons may be competitive, where different teams compete to produce a minimal viable product that best meets some criteria, or cooperative, where all participants work together to build different aspects of a technological system. Unless there is an overarching methodology that structures a partnership in a way that gives real agenda-setting and decision-making power to the civillians involved (for example, the Bristol Approach~\cite{balestrini2017city}), hackathons can struggle to overcome a transactional logic: what is sought is a deliverable, often a technology, and is expected to serve a purpose that is strategically framed by the sponsoring entity~\cite{irani2015hackathons,johnson2014civic}.

As Irani notes, ``Hackathons sometimes produce technologies, and they always, however, produce subjects''~\cite[p.800]{irani2015hackathons}. In other words, civic engagement processes propose and enact specific conceptions political subjectivity and civic performance~\cite{irani2015hackathons,crabu2018bottom}. While at least on the surface, hacknights resemble hackathons, our analysis thus far allows us to demonstrate a crucial difference: an inversion of priorities, placing process over product and relationship over transaction, community over technology. 

This inversion of the hackathon approach was specifically mentioned by participants in interviews. In particular, Gabe one of CTTO's founding members, specifically contrasted hacknights and hackathons:

\begin{quote}
   ``There had been various kinds of civic hackathons in the city before, and those were cool, but we also were tired of hackathons, and so I remember\dots positioning this,\dots~ right at the beginning or soon after,\dots~ as sort of the opposite of a hackathon, where its not about working all through the night to make a thing that's going to suddenly have a ton of impact. It's actually recognizing that the kinds of challenges that we have, civic challenges, are complex challenges, and complex challenges can't be solved in a weekend. They're solved through a real persistent immersion in the problem and bringing different kinds of people together, and trying, and prototyping and trying stuff out and iterating, and that a weekly hacknight is a much better format for doing that.'' (Gabe S)
\end{quote}

This regularity of the hacknights, their temporal embedding into the lives of participants, is seen as essential to making sure they remain casual: a participant might miss a week, a project may stop meeting, but the hacknights will continue. 

One founding member, Bianca, also makes it clear that hacknights try to recuperate what is valuable about the hackathon engagements, by focussing on relationships rather than technologies:

\begin{quote}
    ``I am a frequent defender of hackathons, \ldots in some contexts,\ldots~ they do have a very product development driven intent, but from my experience\ldots~ I found that these kinds of events where you bring people together, it was actually about the relationships and knowledge transfer, and having like, civil servants come out and share things with residents, who were interested in technology, and to learn together\ldots So, I think it's much more a social/political kind of undertaking\ldots [and the] intangible outcomes of these things are much different than the `product' that is created.'' (Bianca W)
\end{quote}

\subsection{Imagining A Civic platform}
Leveraging the ideas of Lefebvre and de Certeau, Korn and Voida describe political experience, or civic engagement as problematically distant from everyday life, which is for the most part depoliticised. In their account, civic engagement, and thus experiences of political agency, are confined ``privileged moments\dots special occasions or punctuated feedback cycles on public servants and service provision''~\cite{korn2015creating}. Korn and Voida's argument is similar to our own: these privileged moments are constructed strategically by experts who are responding to institutional needs. They constrain democratic input to what is `needed' or `deemed appropriate'~\cite{korn2015creating}, which takes the structure of a transaction. 

Exploring how this paradigm of civic engagement can be overcome, and how civic engagement can be infrastructured through design activity, Korn and Voida discuss paradigms of participation that are varyingly based on consensus or contesting, embedded in everyday life or centered on privileged moments. Although Korn and Voida are specifically interested in how technological artifacts embody participatory paradigms, we can also take hacknights and a form on technology, and use their terms to articulate two ways that hacknights reconfigure civic engagement.

Participation in CTTO is form of what Korn and Voida call ``Situated Participation'', leveraging technologies like messaging, scheduling, and shared drives to embed civic activity within temporal, social, and spatial aspects of everyday life~\cite{korn2015creating}. CTTO hacknights take considerations of public life out of the privileged moments of government-led processes and into a space of active participation where interest, enjoyment, and personal growth are prioritized and integrated as part of civic activity. At CTTO, civic activity is not constrained by the same focus on issues that characterizes publics, but rather is located in an ethos of cooperation and a network of peers that can support the emergence of publics. Hacknights are not purely instrumental happenings intended to produce applications, but serve as a continual site of circulation that itself moves across the city. 

As an ongoing and community supported practise, civic tech probes the boundaries of bureaucratic processes, seeking ways to make them more porous and interactive. It does this through a focus on the work of public servants, the drafting of policy and the creation of public technologies, rather than adopting a partisan stance in support of elected officials. It moves its participants, public servants and civilians, closer to each other, and into positions of cooperation and experimentation. This challenges approaches to governance that, even when they engage in participatory or user centred design, configure publics as users of services for which they are the provider. Hacknights are a place for what Erhardt Graeff calls ``civic learning'', by which professionals (especially technologists) come to understand the stakes of governance and the politics of technology, such as through interactions with public servants and community members~\cite{graeff2014crowdsourcing}. Enabling these types of technical and social literacies is crucial to realizing meaningful participation in technologically sophisticated democracies, and the types of practical and experimental encounters that workshops and hacknights offer are an effective means of facilitating these literacies and building agency~\cite{peer2019workshops}.
Hacknights not only equip the public with the technological and policy literacies they need to engage with governance, but acquaint public servants with the value and expertise that these collaborations can bring. The conception of citizenship enacted at the hacknights are active, rooted in personal action and experience, and centred in a community, rather than based on (material) interests of the individual, and the notion that the business of the state should interact as little as possible with everyday life. 

In these ways, CTTO serves as an example of how diverse groups of participants form communities that revolve around civic engagement and a commitment to interacting with governments. If civic cooperation, and even participation in public decision-making, are effectively catalyzed in group settings, then infrastructuring the connections between these groups and the state provides a tempting vector for research and praxis. CTTO exemplifies how community coalitions can build the capacities required for sociotechnical citizenship and seek out engagements with governance. 

\subsection{Summary and Contributions}
Using examples of projects from CTTO's history and practises, we have argued for the importance of considering civic technology as about, but not reducible to technological artifice. To realize the democratic potentials of civic technology, it is just as important to examine how local context affects \textit{how people relate} to the work of governance and how the practises of civic technology groups can prepare and enhance the capacities and opportunities for people \textit{as} political subjects. 

Prior CSCW work too has emphasized the importance of understanding the relationships between government, civilian and organizational actors surrounding civic tech~\cite{vlachokyriakos2016digital,asad2017creating,corbett2018problem,dickinson2019cavalry}, the question of research methods~\cite{aragon2020civic}, and the importance of context. The research presented directly addresses these questions. In terms of methods, because of the highly social nature of communities like CTTO, action research emerges as an appropriate way of studying how civic technology helps to connect practitioners to the state. It is important to note the importance of long-term involvement in community organizing work to this research method. There are ample opportunities for community-oriented forms of research, joining existing grassroots communities to help them do their work. Researchers are valuable to these communities because their expertise as academic: they can help people think through issues, they often have subject matter expertise or pedagogical skill that can help others learn, they can draw on institutional resources, and they are able to mediate between experts and laypeople. Their academic authority can provide weight to their concerns and credence to their perspectives, and their research time can be simultaneously beneficial to the community. It is important to seek outcomes less in tangible products and more in the community itself.

At CTTO, the focus lies not on the technologies that are imagined or created, but on the practises and people involved in the work. Importantly, by following this shift of focus in our study, the de-centering of technologies has helped us focus on the relations enacted by local civic tech practise. CTTO shows how civic tech can transcend the generally transactional frame usually implicit within studies of civic technologies. Supporting the autonomous and social nature of these communities, and finding ways to solidify relations between states and civics, is essential to building the conditions for democratic governance. Participants recognize the democratic potential of this shift:

\begin{quote}
``I get really excited about spaces like that, where a place of real power, government, is adjacent, because then they get to be involved in that learning, too, \ldots{} how power works, by seeing it move and change.'' (Patrick C)
\end{quote}

\section{Conclusions}
\label{conc}

Di Salvo et al. note that when brought to bear on matters of concern, through design centred ``activities of re-imagination, designers and participants are engaging in prototyping new social, economic, and political arrangements''~\cite[p.2405]{disalvo2014making}. If civics are sociotechnical, then people will become the types of political subjects that our democracies ask and allow them to be. Providing opportunities to engage with democratic processes is thus a necessary first step in improving the quality of democratic participation. Pateman elegantly summarizes: ``we do learn to participate by participating, and\ldots{} feelings of political efficacy are more likely to be developed in a participatory environment''~\cite[p.105]{pateman1970participation}.

CTTO operates as experimental site for civic engagement, demonstrating an expanded set of interactions and work objects that can form the basis for participation in the work of public technology design and decision-making. The hacknights create a contact zone between a diverse range of public expertise and civil servants, create opportunities for collaborative projects, and operate as a key site for ``civic learning'' ~\cite{graeff2014crowdsourcing}. Civic learning is a necessary condition for realizing the potential benefits, both epistemic and legitimating, of participatory democracy. The exchanges between public servants and publics are reciprocal. Publics provide valuable perspectives and ideas, while also potentially legitimating ongoing work. In turn, public servants can help publics navigate the intricacies of their bureaucracies, and are effective stewards of institutional knowledge. Support from public institutions also helps legitimate the work of the community.

As a community, CTTO is held together by a unique set of performed rituals that  creates an intermediate space between public life and the public sector. While the value of the hacknights is not easily expressed in terms of measurement or productive capacity, this absence of productivity pressure allows participants to take agency, to decide what should be done, and to enact it themselves without fear of failure. Organizers set the stage for civic action within an entity that is unlike an issue-oriented organization or a formal advisory group. Flexible, supportive and configurable, CTTO creates a platform for civic engagement that seeks to demystify both government and technology. 

Throughout these discussions, it is clear that understanding civic tech requires a lens beyond the mere analysis of the production of technical artifacts, but rather signifies a practise or process for making technologies that is social and participatory. Here, our study corroborates and extends the findings of earlier studies, especially in emerging areas of research such as Digital Civics. 

Nonetheless, many questions remain. There is opportunity for research whenever civic groups come into regular contact with public servants, especially when the research is based in an existing community, rather than when governments try to elicit participation. We were privileged to have access to such a longstanding and interesting community in our locale, and researchers should, whenever possible, seek out partnerships with local groups. Researchers and academics can draw on their own expertise in analysis, organising and technology to put themselves in service of community goals, and leverage university resources to support them as well.  

Studying civic tech can shed new light on the relationships between communities vs. institutions~\cite{cooke2012the}. One of the ways that civic tech practise can reconfigure democratic relationships is by bringing civilians and public servants into productive, albeit informal contact based around cooperative practise. How can civic tech make the expertise of public servants available to publics? How are these relationships infrastructured, and made more durable? Can this contact help communities develop their own solutions? Have technologies been used to accomplish this? Further work could also focus more specifically on whether subverting formal processes of engagement or consultation in neutral contact zones can be just as helpful to public servants as it can be to civilian groups. 

It would also be useful to ask when and how can these relationships informal collaborations lead for more formalized partnerships, or about local or institutional obstacles to realizing these goals. Seeking to climb the ladder of participation, we can ask what organizational forms leave communities in control of civic tech and policy? Research in this field would be greatly advanced by rich and historically contextualized accounts of instances where civilian groups, civic tech or not, were able to gain and claim power in technology design or policymaking processes. 

Further research can also continue to bring nuance and critique to these kinds of democratic participation. Apart from continuing to deploy frameworks like the ladder of participation and pseudo-participation, how can we as researchers begin to navigate the sometimes obscure boundary between empowering people and responsibilizing them for what should be paid labour? What kind of `trust work' and which forms of civic tech can support trust of communities in public servants in situations when elected representatives cannot be trusted (for example, if they too are not accountable to engagement results)? 

What would it mean to embrace this more porous, collaborative, and relational model of public service within government? By virtue of its informal nature, there are severe challenges to the typically risk-averse and hierarchical nature of public service work. Nonetheless, bringing civilians and public servants into sustained contact helps to demonstrate that realizing goals of transparency and open government, much less participation, cannot be accomplished by merely making changes to bureaucratic processes or by providing more information-- it must be done through cultivating relationships. Groups like CTTO are just the kind of entities that are capable of organizing citizens on their own terms and providing a venue for governments to engage. The work of these groups is crucial to helping governments address their methodological and organizational challenges of participatory democracy, and it exhorts us researchers to focus on how this democratic progress is as much social as it is technological.

\begin{acks}
Enormous gratitude goes to the many participants from Civic Tech Toronto who contributed to this research in myriad ways. Thank you to Matt Ratto, whose ideas about transactive, performative, and enactive object relations influenced our thinking on the kinds of relationships supported at Civic Tech Toronto. 

This research was partially supported by Natural Sciences and Engineering Research Council through RGPIN-2016-06640, the Canada Foundation for Innovation, Onatrio Research Fund, and the Social Science and Humanities Research Council through the Canadian Graduate Scholarship 752-2018-2511.
% Matt's contributions mentioned here?
\end{acks}
\bibliographystyle{ACM-Reference-Format}
\bibliography{cscw.bib}
\received{January 2022}
\received[revised]{July 2022}
\received[accepted]{November 2022}

%%TODO pointers CB found quickly:
% https://dl.acm.org/doi/abs/10.1145/224019.224035
% Engaging with Practices: Design Case Studies
% as a Research Framework in CSCW. https://www.wineme.uni-siegen.de/paper/2011/p505-wulf.pdf
% also maybe useful: https://www.slideshare.net/LuiginaCiolfi/qualitative-methods-in-cscw-research
% https://research.ibm.com/publications/mapping-the-how-of-collaborative-action-research-methods-for-studying-contemporary-sociotechnical-processes too

\end{document}